\documentclass[10pt]{article} 
\usepackage[preprint]{tmlr}

\usepackage{amsmath,amsfonts,bm}

\def\eqref#1{equation~\ref{#1}}

\def\1{\bm{1}}

\DeclareMathAlphabet{\mathsfit}{\encodingdefault}{\sfdefault}{m}{sl}
\SetMathAlphabet{\mathsfit}{bold}{\encodingdefault}{\sfdefault}{bx}{n}

\usepackage{hyperref}
\usepackage[table,xcdraw]{xcolor}
\usepackage{multirow}
\usepackage{url}            
\usepackage{booktabs}       
\usepackage{amsfonts}       
\usepackage{nicefrac}       
\usepackage{microtype}      
\usepackage{amsmath,graphicx}
\usepackage{subcaption}
\usepackage{amssymb}
\usepackage{xspace}
\usepackage{enumitem}
\usepackage[accsupp]{axessibility} 
\usepackage{breqn}
\usepackage{minted}
\usepackage{algorithmic}
\usepackage{algorithm2e}
\usepackage{bbm}

\title{Beyond Reconstruction: What EEG-to-Video Decoding Actually Recovers\thanks{Project page at \url{https://prajwalsingh.github.io/EEGVid/} and code at \url{https://github.com/prajwalsingh/EEGVid}.}}

\author{\name Prajwal Singh, Anupam Sharma, Pankaj Pandey, Krishna Miyapuram, Shanmuganathan Raman \\
      \addr CVIG Lab and BraIn Lab \\
      \addr IIT Gandhinagar \\
      \addr Email: \texttt{\{singh\_prajwal, sharmaanupam, pankaj.p, kprasad, shanmuga\}@iitgn.ac.in}}

\def\month{MM}  
\def\year{YYYY} 
\def\openreview{\url{https://openreview.net/forum?id=XXXX}} 

\usepackage{multirow}

\usepackage{listings}
\usepackage{xcolor}

\definecolor{codegreen}{rgb}{0,0.6,0}
\definecolor{codegray}{rgb}{0.5,0.5,0.5}
\definecolor{codepurple}{rgb}{0.58,0,0.82}
\definecolor{backcolour}{rgb}{0.95,0.95,0.92}

\lstdefinestyle{mystyle}{
    backgroundcolor=\color{backcolour},   
    commentstyle=\color{codegreen},
    keywordstyle=\color{magenta},
    numberstyle=\tiny\color{codegray},
    stringstyle=\color{codepurple},
    basicstyle=\ttfamily\footnotesize,
    breakatwhitespace=false,         
    breaklines=true,                 
    captionpos=b,                    
    keepspaces=true,                 
    numbers=left,                    
    numbersep=5pt,                  
    showspaces=false,                
    showstringspaces=false,
    showtabs=false,                  
    tabsize=2
}

\begin{document}
\maketitle

\begin{abstract}\vspace{-1.0em}
Reconstructing \emph{dynamic visual stimuli} from EEG recordings is challenging due to the noisy, non-stationary nature of EEG signals and the limited availability of EEG-video datasets. We present EEGVid, a framework that learns EEG representations using triplet loss and reconstructs dynamic videos with a temporally conditioned GAN. We study what these representations encode and how this information supports generation. First, visual representations retain emotional structure, while emotion-based supervision does not preserve the same fine-grained visual information. Second, triplet learning shifts EEG features away from subject-specific structure toward stimulus-related information. Third, analysis across brain regions, hemispheres, and time reveals consistent differences in visual and emotional encoding, with temporal regions contributing strongly across tasks. Finally, we evaluate video generation using three controlled diagnostics. The learned encoder separates unseen video classes better than a randomly initialized encoder, although the cross-subject gain is small, while mismatched EEG conditioning shifts generated content toward the substituted stimulus, showing that the generator actively uses EEG as a content signal. A ground-truth class label yields stronger reconstruction metrics, partly because the label is a cleaner conditioning target. Together, these results show that EEG exhibits a consistent visual and emotional structure that can support dynamic video generation, whereas current reconstruction primarily reflects coarse stimulus-level information rather than fine-grained, segment-specific decoding.
\end{abstract}    
\section{Introduction}

Understanding how the brain encodes dynamic visual experiences remains a central challenge in neuroscience and computer vision. Unlike static perception, natural vision unfolds as a continuous stream that engages multiple cognitive and affective processes \citep{wandell1999computational,fish2021philosophy,kosslyn1991cognitive}. Recent advances in multimodal representation learning have enabled neural decoding, mainly from fMRI, by aligning neural signals with pre-trained image or text models. However, it remains unclear how much information \emph{EEG}, which has millisecond temporal resolution but low spatial precision, carries about dynamic visual stimuli when studied without other modalities such as images or text.


\begin{figure*}[!t]
\centering
\begin{minipage}[b]{0.52\textwidth}
  \centering
  \includegraphics[width=\linewidth]{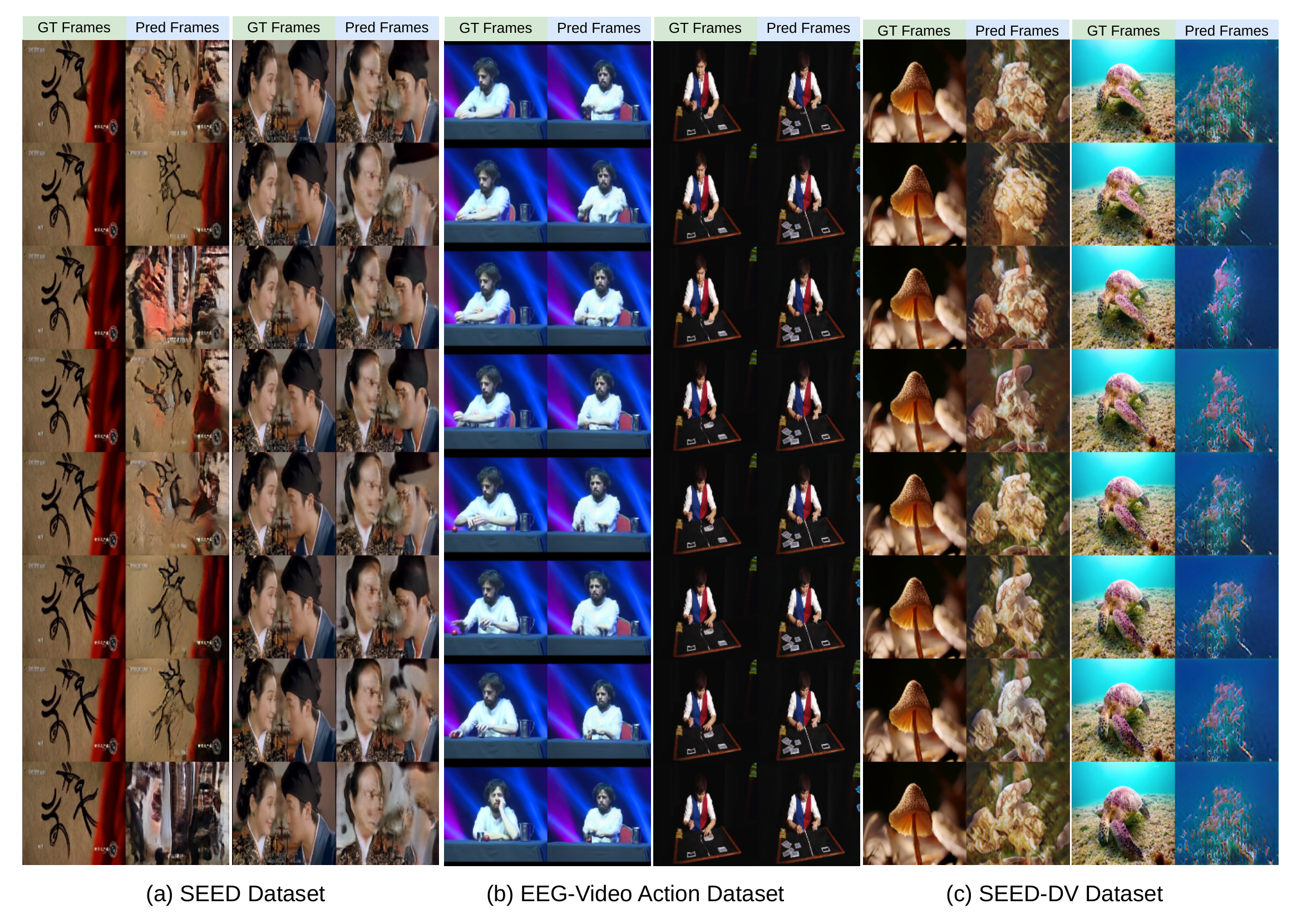}\vspace{-0.5em}
  \centerline{\small (a)}
\end{minipage}\hfill
\begin{minipage}[b]{0.19\textwidth}
  \centering
  \includegraphics[width=\linewidth]{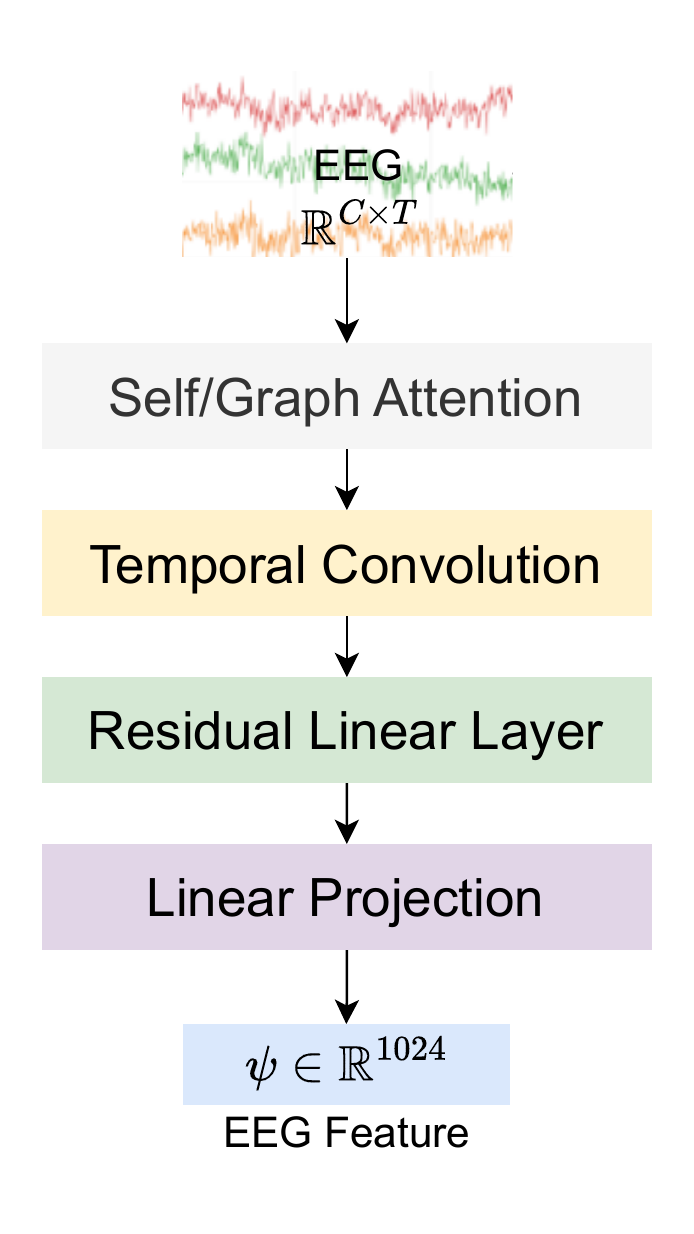}\vspace{-0.5em}
  \centerline{\small (b)}
\end{minipage}\hfill
\begin{minipage}[b]{0.26\textwidth}
  \centering
  \includegraphics[width=\linewidth]{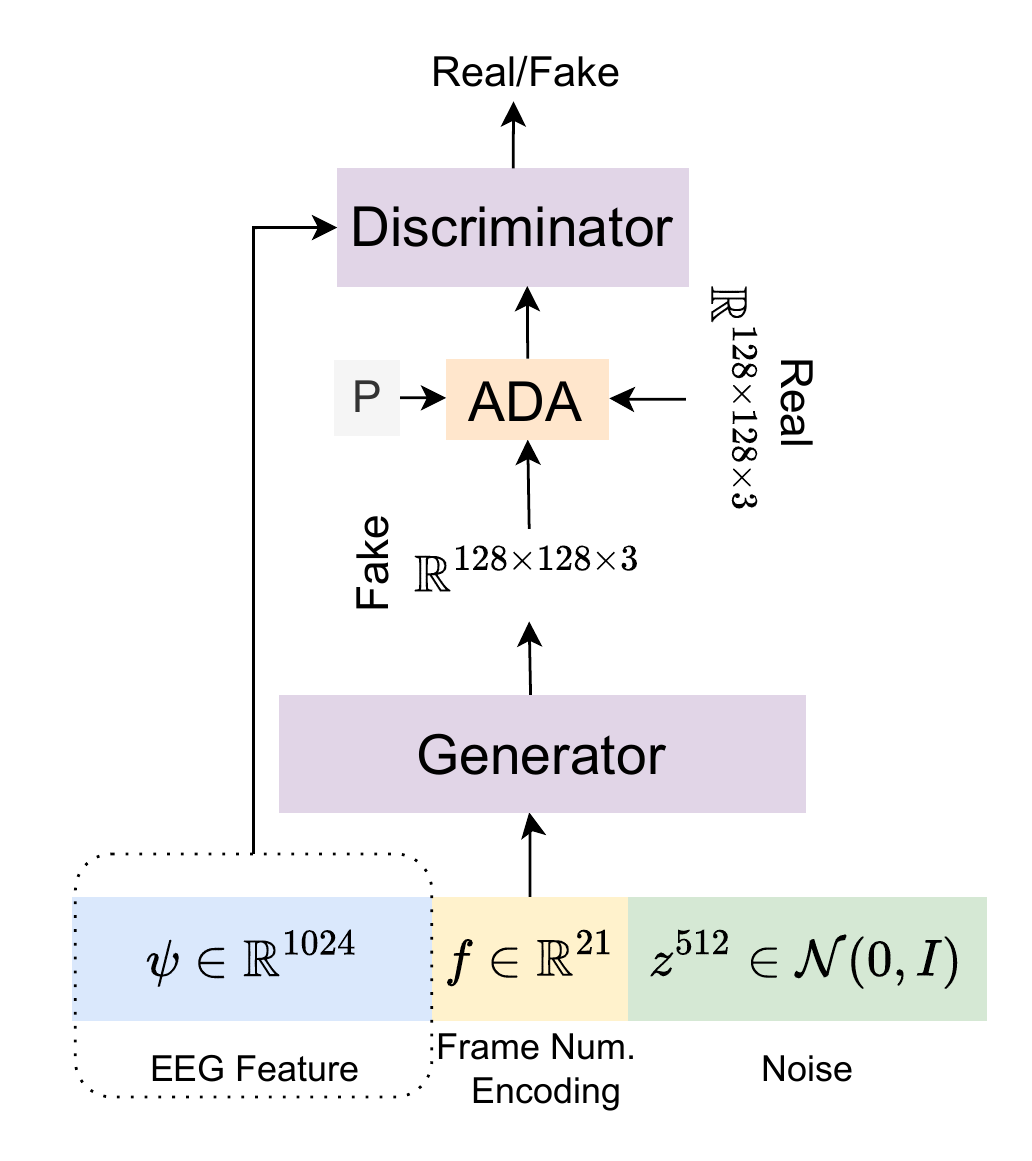}\vspace{-0.5em}
  \centerline{\small (c)}
\end{minipage}\vspace{-0.8em}
\caption{\textbf{Overview of EEGVid.} (a) Video frames synthesized from EEG for clips shown to participants in the SEED~\protect\citep{duan2013differential,zheng2015investigating}, EEG-Video Action~\protect\citep{yao2024identifying}, and SEED-DV~\protect\citep{eeg2video} datasets. Each clip has eight frames. The EEG provides semantic guidance, and the dataset-specific GAN prior provides fine spatial detail. (b) A modified NICE EEG encoder~\protect\citep{song2023decoding}, trained with triplet loss, maps EEG to a $1024$-d representation. (c) A StyleGAN-ADA~\protect\citep{karras2020training} generator synthesizes each frame from this representation, concatenated with a positional encoding of the frame index.}
\label{fig:teaser}
\label{fig:eeg2vid_archit}
\vspace{-0.5em}
\end{figure*}

To address this question, we use EEG-video datasets that include both emotional and non-emotional stimuli and learn discriminative EEG representations via triplet-based contrastive learning. We then use these representations in a temporally conditioned GAN to generate video clips that visualize the information carried by the EEG.

A central challenge in this setting is evaluation. Strong reconstruction quality on a fixed training distribution can have two very different explanations. It may reflect genuine segment-specific EEG decoding, or a generator that has memorized a small set of training stimuli and is guided only by the video's coarse category. Reconstruction metrics alone cannot distinguish these cases, and the actual result is likely a combination of both. Without controls to separate these effects, a pipeline can achieve strong reconstruction scores even when the EEG encoder carries little more than category information, with the generator's memorized prior providing the remaining detail. Addressing this evaluation gap, while also understanding what EEG itself encodes, is a central goal of this work.

We therefore investigate the following research questions:

\begin{itemize}
    \item \textbf{RQ1: How are visual and emotional features related?} In the SEED dataset, emotions are induced through video content. We test whether visual features also carry affective information. Our results show that emotional information is largely content-driven, with visual representations showing visible affective structure in t-SNE.
    
    \item \textbf{RQ2: Do learned visual features retain subject-specific information?} While untrained features cluster by subject, learned features emphasize stimulus information and reduce subject-related bias.
    
    \item \textbf{RQ3: How do different brain regions contribute to encoding?} We find that the temporal lobe, with only six channels, gives the highest accuracy in five of the six SEED and EEG-Video Action settings, the left hemisphere dominates for video classification and the right hemisphere for emotion, and, when training and test subjects overlap, frontal channels outperform posterior channels in both tasks.
    
    \item \textbf{RQ4: Can dynamic videos be reconstructed from EEG, and how much of this reconstruction reflects genuine segment-specific decoding?} By conditioning a GAN on EEG features and temporal indices, we show that generated frames align semantically with the videos participants observed. We then introduce three controlled diagnostics, i.e., a held-out-stimulus generalization test for the encoder, a mismatched-conditioning control for the generator, and a class-conditional baseline that compares EEG with a ground-truth label. These tests identify which aspects of the reconstruction the EEG drives.
\end{itemize}

By addressing these questions, we demonstrate dynamic video reconstruction from EEG alone and provide quantitative insights into \emph{what} EEG encodes about visual and emotional dynamics across regions, hemispheres, and time. We also provide a controlled evaluation of how much of the reconstruction reflects genuine segment-specific decoding rather than coarse category cues or generator memorization.
\section{Related Work}
\label{sec:relatedwork}

\vspace{1mm} \noindent
\textbf{Decoding Static Perception.} Reconstructing visual images from brain activity has been widely studied using fMRI signals. Early work used linear and Bayesian models to map neural responses to visual stimuli~\citep{34,35,36,37}. \citet{34} combined multiscale decoders to reconstruct basic patterns, while \citet{35} used a Bayesian framework for natural images. \citet{36} reconstructed color from responses in the early visual cortex, and \citet{37} extended linear decoding methods.

Recent approaches have adopted deep learning for neural response mapping. \citet{38} and \citet{39} used self-supervised learning for natural image reconstruction. \citet{40} proposed “Mind Reader” for reconstructing complex images. GAN-based methods~\citep{41,42} further improved semantic realism, building on earlier approaches~\citep{43,44}.

More recently, latent diffusion models have enabled high-resolution visual reconstruction. \citet{20}, \citet{21}, and \citet{contrast_atten_diffuse} conditioned diffusion models on brain activity to generate detailed images. \citet{23}, \citet{24}, and \citet{25} introduced semantic and controllable conditioning to improve reconstruction quality and interpretability. \citet{26} demonstrated latent diffusion for natural scene reconstruction, while \citet{27} trained shared-subject models and \citet{28} decoded MEG in real time. These advances build on diffusion models~\citep{47,45,46}, which provide effective tools for visual reconstruction from brain signals.

\vspace{1mm} \noindent
\textbf{Decoding Dynamic Perception.} Decoding dynamic perception from brain signals has progressed from linear models to deep generative approaches. Videos activate broader cortical regions than static images~\citep{buccino2001action,schultz2009natural}, making video reconstruction a more challenging setting. Several studies have explored the reconstruction of continuous video stimuli from fMRI~\citep{wen_neural_encoding_decoding,penny_for_your_thoughts,brain_netflix,neuralflix,fmri_video_gan,cinematic_mindscapes}.

\citet{wen_neural_encoding_decoding} used linear regression to estimate feature maps from fMRI and passed them to a Deconvolutional Neural Network~\citep{de_cnn} for reconstruction. \citet{penny_for_your_thoughts} proposed an encoder-decoder framework with a self-supervised stage that generated synthetic fMRI from video using a frozen encoder before training the decoder to reconstruct frames. NeuroClips~\citep{gong2024neuroclips} guided video diffusion with reconstructed keyframes and low-level perception. Our work focuses on the more challenging but temporally richer setting of \emph{EEG-only} decoding.

To improve data efficiency, \citet{neuralflix} aligned fMRI features with CLIP~\citep{radford2021learning} embeddings and used diffusion for video generation. \citet{brain_netflix} followed a similar approach using Masked Brain Modeling~\citep{21}. \citet{cinematic_mindscapes} introduced Spatiotemporal Attention to better capture spatial structure and account for hemodynamic lag~\citep{buckner1998event}. GAN-based approaches have also been explored~\citep{fmri_video_gan}.

Despite these advances, fMRI is costly and has low temporal resolution. EEG provides a cheaper alternative with much higher temporal resolution. \citet{eeg2video} introduced the SEED-DV dataset and EEG2Video, a Seq2Seq model for reconstructing videos from EEG. However, classification accuracy on this dataset is low for current encoders ($6.20\%$ for $40$ classes with their GLMNet encoder in our reproduction, and $2.9\%$ for ours, chance $2.5\%$, Appendix~\ref{sec:appendix_seedandseeddv}), which limits both the learned representations and the generated videos. Unlike EEG2Video, our approach uses a GAN-based framework to generate videos directly from EEG features, avoiding cross-modal alignment~\citep{liang2022mind} and dependence on large pre-trained networks.

\vspace{1mm} \noindent \textbf{Concurrent EEG-to-Video Reconstruction.} A growing body of concurrent work pursues higher-fidelity EEG-to-video reconstruction through more elaborate architectures. DynaMind~\citep{dynamind2025} and MindCine~\citep{mindcine2026} align EEG with multimodal semantics and large-scale pretrained models before diffusion-based decoding, EVOKE~\citep{jing2026evoke} decouples visual, semantic, and motion information via implicit neural representations, and EEGMirror~\citep{eegmirror2025} and \citet{han2026cross} target cross-subject generalization through montage-agnostic pretraining and adversarial subject-invariant mapping, respectively. NEED~\citep{need2025} unifies cross-subject and cross-task generalization within a single architecture. These methods report substantial gains over EEG2Video and represent the current state of the art in reconstruction fidelity. Our goal is different. Rather than pursuing higher reconstruction quality, we ask what a given level of reconstruction quality actually reveals about the underlying neural signal, a question orthogonal to architectural improvement and, to our knowledge, not addressed by this concurrent work. We therefore build on a comparatively simple triplet-encoder-plus-GAN pipeline and direct our contribution toward the diagnostic controls of Section~\ref{sec:diagnostics}, which apply equally to any of the architectures above.

\section{Method}
\label{sec:method}

In this section, we describe how and why we use a triplet-loss-based feature extraction method \citep{schroff2015facenet} to learn EEG representations. We then describe how we use EEGStyleGAN-ADA \citep{singh2024learning} as a generative probe that turns these representations into video frames.

\subsection{Extracting Features from EEG}
\label{subsec:eeg2feat}

A key step in generating videos from EEG signals is extracting meaningful visual features. We follow established contrastive learning approaches \citep{mishra2021eeg, singh2023eeg2image} to learn representations from EEG data. Specifically, we use triplet learning \citep{schroff2015facenet} with an anchor ($x^{a}$), positive ($x^{p}$), and negative ($x^{n}$) sample. As shown in Eqn.~\ref{eqn:1}, the triplet loss pulls the positive feature closer to the anchor while pushing the negative feature away. $\delta$ denotes the margin between the positive ($x^{p}$) and negative ($x^{n}$) pairs.\vspace{-0.5em}
\begin{align}
    \min_{\theta}\mathbb{E}\big[ \|f_{\theta}(x^{a}) - f_{\theta}(x^{p})\|_{2}^{2} - \|f_{\theta}(x^{a}) - f_{\theta}(x^{n})\|_{2}^{2} + \delta \big]_{+}
    \label{eqn:1}
\end{align}

\vspace{1mm} \noindent
\textbf{Why use the Triplet Method?} Recent works \citep{neuralflix,brain_netflix,cinematic_mindscapes} use CLIP \citep{radford2021learning} to learn brain representations by aligning them with pre-trained image or text features. In contrast, we use triplet loss to learn the representation directly from the EEG. It uses labels to form positive and negative pairs, encouraging the network to learn discriminative patterns in the EEG signals. This keeps the learned representation free of structure imported from another modality, so any structure we later find in it comes from the EEG. Since the EEG signals in our setting are responses to video stimuli, triplet learning also encourages features that capture information related to the observed videos.

\begin{figure*}[!t]
\centering
\begin{minipage}[b]{0.245\linewidth}
  \centering
  \centerline{\includegraphics[width=\linewidth]{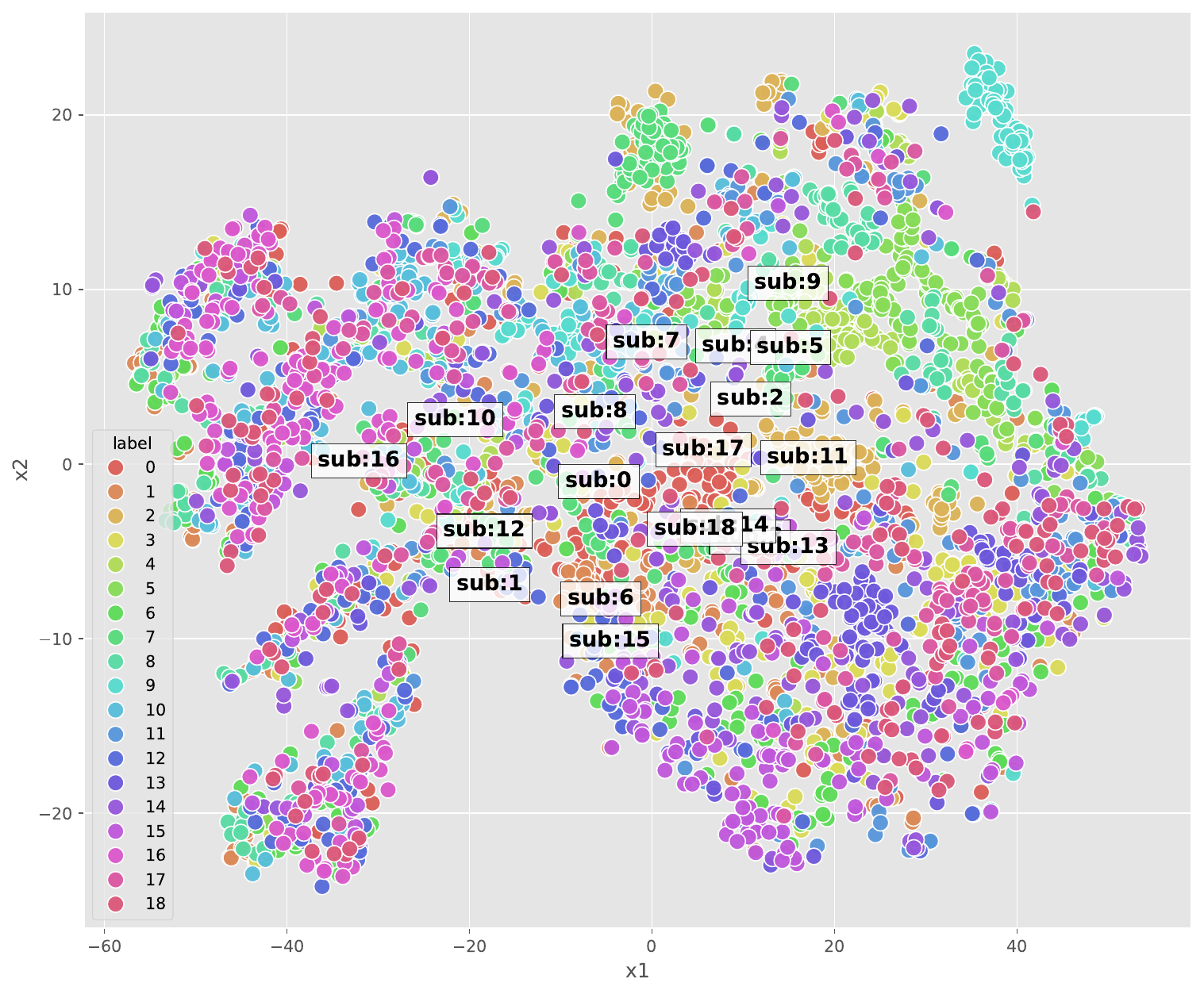}}\vspace{-0.2em}
  \centerline{\small (a)}\medskip
  \vspace{-4mm}
\end{minipage}
\begin{minipage}[b]{.245\linewidth}
  \centering
  \centerline{\includegraphics[width=\linewidth]{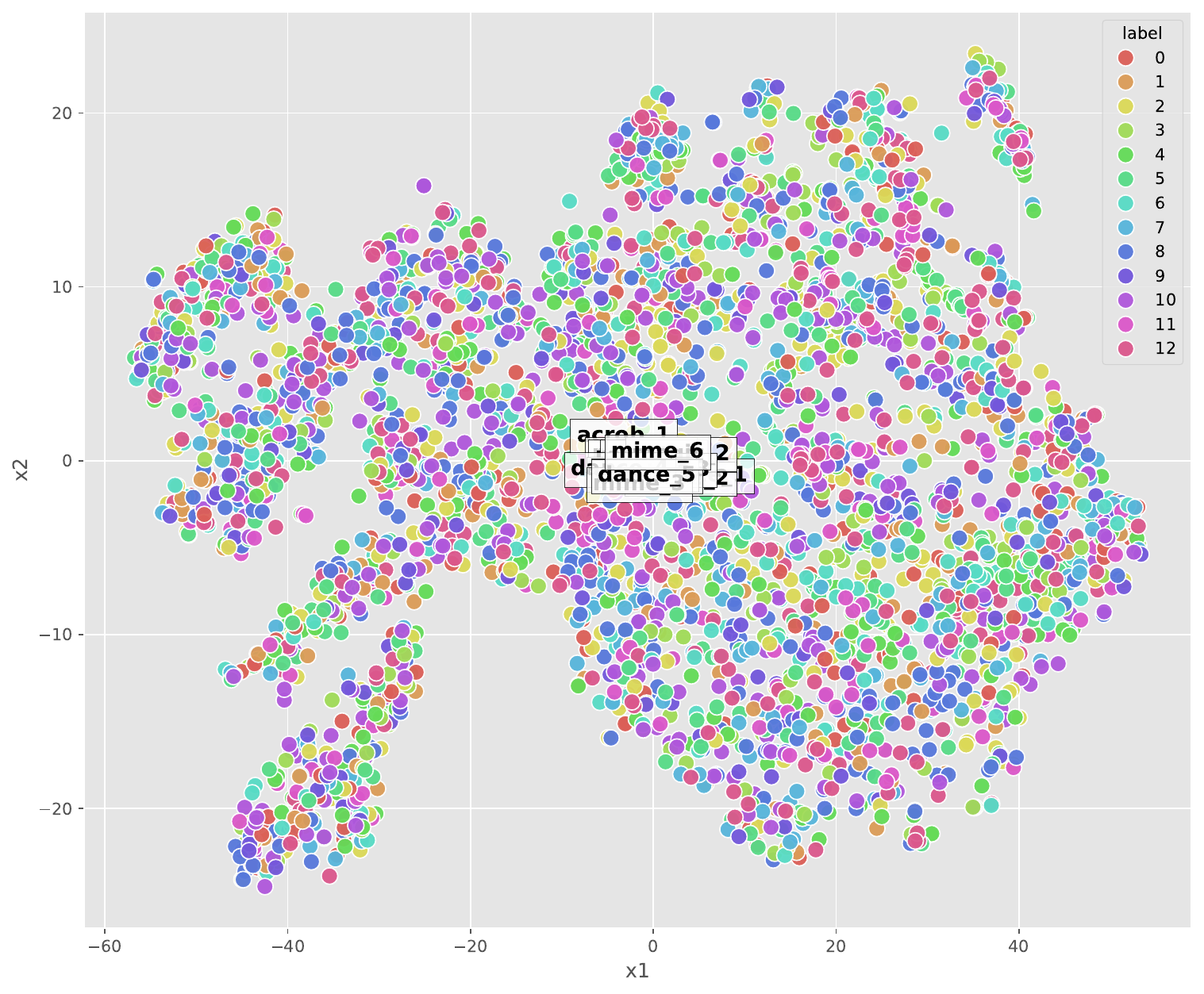}}\vspace{-0.2em}
  \centerline{\small (b)}\medskip
 \vspace{-4mm}
\end{minipage}
\begin{minipage}[b]{0.245\linewidth}
  \centering
  \centerline{\includegraphics[width=\linewidth]{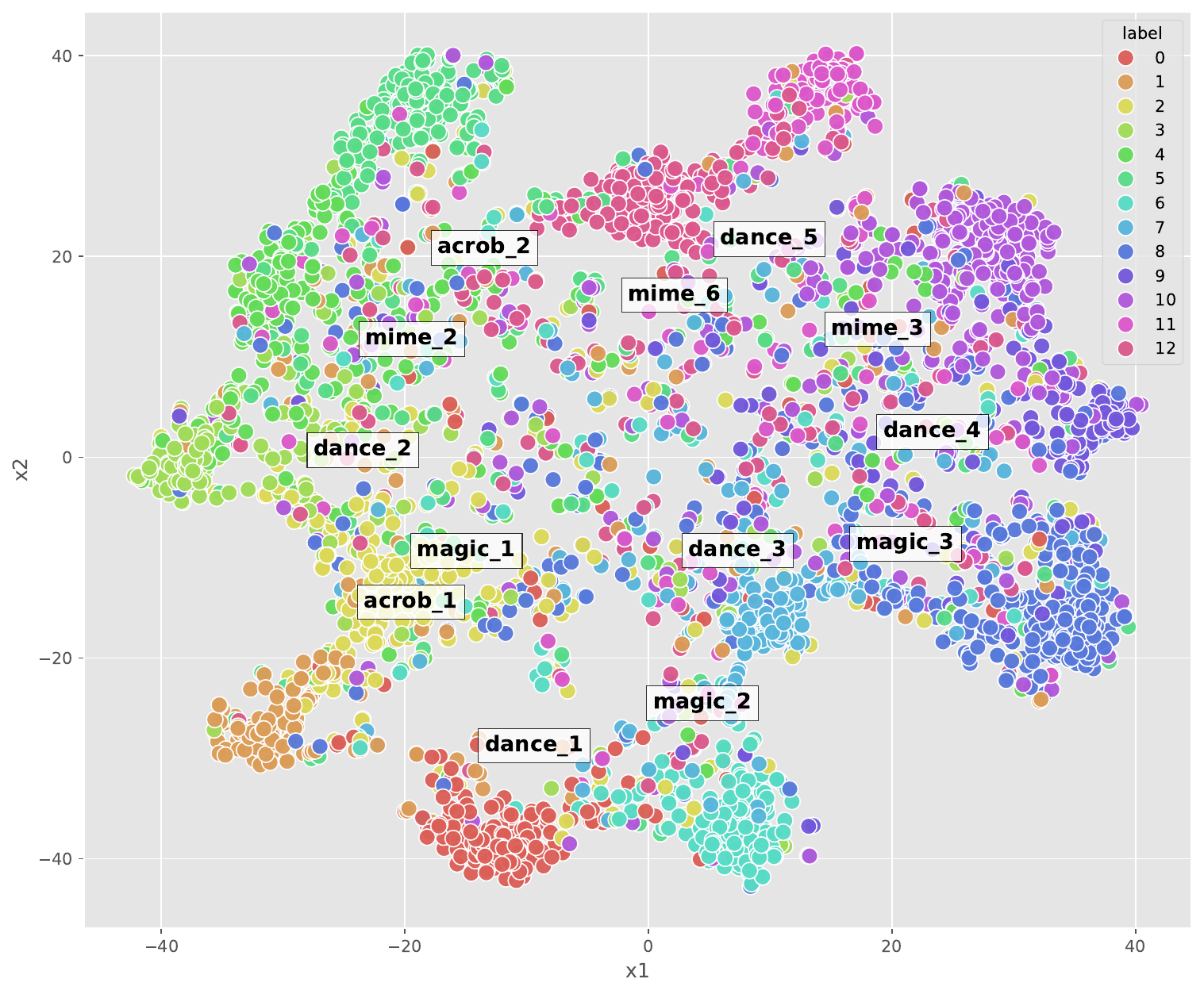}}\vspace{-0.2em}
  \centerline{\small (c)}\medskip
 \vspace{-4mm}
\end{minipage}
\begin{minipage}[b]{0.245\linewidth}
  \centering
  \centerline{\includegraphics[width=\linewidth]{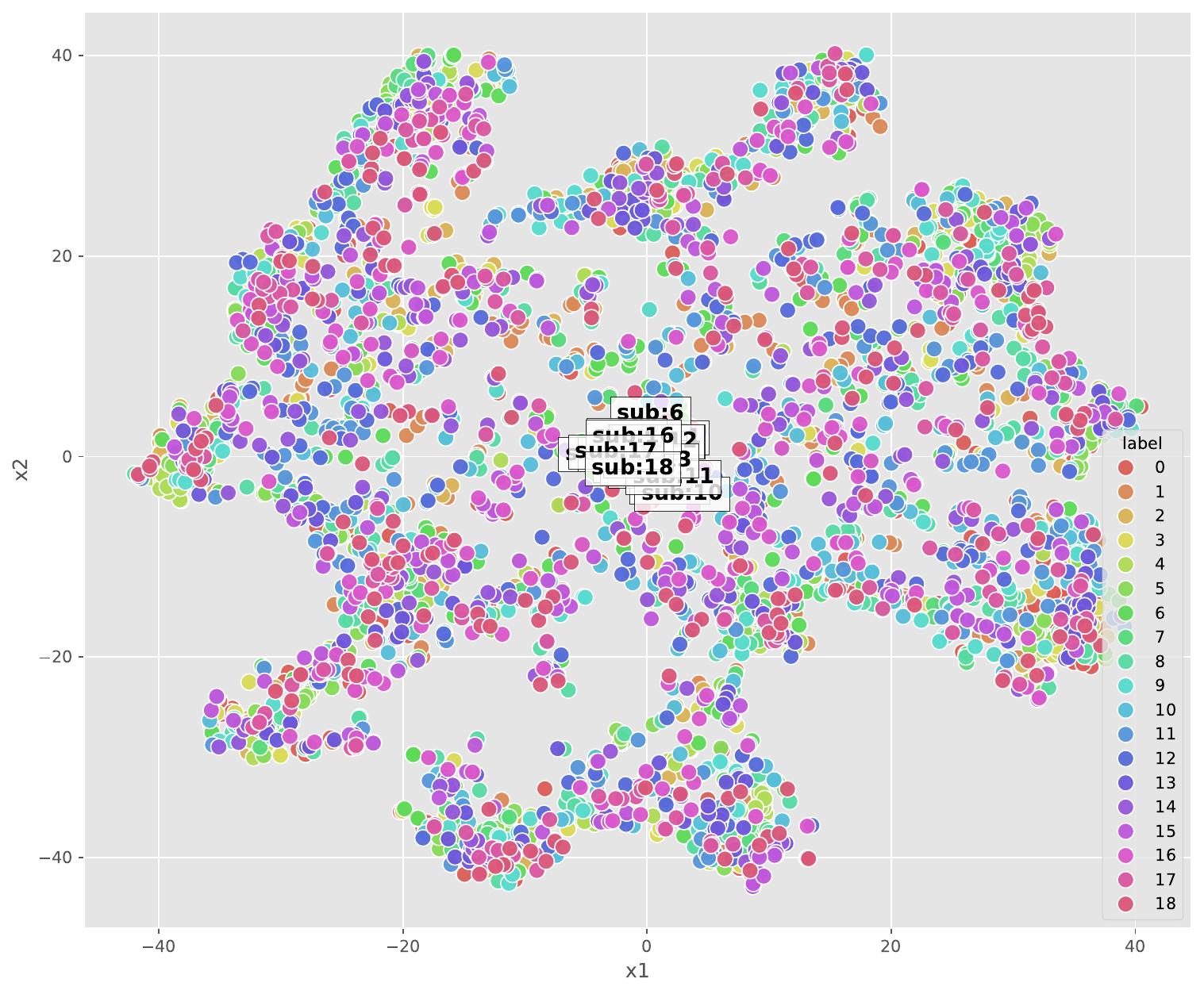}}\vspace{-0.2em}
  \centerline{\small (d)}\medskip
 \vspace{-4mm}
\end{minipage}\vspace{-0.5em}
\caption{\textbf{Effect of triplet loss} (Zoom in for detail). EEG embeddings via t-SNE under different training conditions on the EEG-Video Action dataset. (a-b) Randomly initialized encoder: features cluster strongly by subject ID (19 subjects) but collapse into a single cluster when grouped by video stimulus (13 videos), showing little video-related information before training. (c-d) After triplet-loss training on video labels, the pattern reverses: embeddings cluster distinctly by video stimulus ($\sim$70\% k-means accuracy) while subject ID clustering collapses. Triplet loss thus reorients EEG representations away from subject identity and toward stimulus-related video information.}
\label{fig:tsne_plots_single_shot}
\vspace{-1.2em}
\end{figure*}

Another important part of EEG representation learning is the encoder architecture. We use a modified NICE EEG \citep{song2023decoding} network. As shown in Figure~\ref{fig:eeg2vid_archit}(b), it consists of a graph-attention layer (GAT) \citep{velivckovic2017graph}, followed by temporal convolution and a linear layer that produces a $1024$-dimensional feature $\psi$. The GAT layer combines information from all EEG channels into a shared representation while weighting them according to their relevance.

\vspace{1mm} \noindent
\textbf{Triplet Pairs.} We construct triplet pairs as follows. a) EEG segments from the same video clip form a positive pair for video classification, while EEG segments from the same emotion class form a positive pair for emotion classification. b) EEG segments from different videos or emotion classes form negative pairs. Because positives can come from different subjects, this encourages the network to learn stimulus-related features while reducing subject-specific structure.

\subsection{Generative Probing for Visualizing Decoded Representations}
\label{subsec:feat2video}

To synthesize video frames from the learned EEG representation, we follow \citet{singh2024learning} and use the StyleGAN-ADA \citep{karras2020training} architecture shown in Figure~\ref{fig:eeg2vid_archit}(c). We train EEGStyleGAN-ADA \citep{singh2024learning} from scratch at $128 \times 128$ resolution and modify its conditioning input because the original architecture is designed for image synthesis. We concatenate the EEG feature with a positional encoding of the frame number ($f$), which represents the temporal position of the frame in the video, following \citet{pumarola2021d}. We use sinusoidal positional encoding with frequencies $2^{k}$, as in D-NeRF, for the frame number, with $10$ frequencies ($21$-d) for SEED and SEED-DV and $2$ frequencies ($5$-d) for EEG-Video Action. All EEG-Video Action results, including the diagnostics in Section~\ref{sec:diagnostics}, use the same $5$-d model. This conditioning allows the same EEG feature to generate different frames at different temporal positions. We also modify the generator loss \citep{karras2020training} by adding an $L_{1}$ loss against the ground truth frame, with weights $\lambda_{1}=0.5$ and $\lambda_{2}=5.0$.\vspace{-0.5em}
\begin{equation}
    \mathcal{L}_{total}^{Gen} = \lambda_{1} \mathcal{L}_{Gen} + \lambda_{2} \mathcal{L}_{1}(Gen, Real)
\end{equation}

The positional encoding is what makes the frames move. Without it, the optical-flow score (OFS) on EEG-Video Action is $0.276$, below the $0.479$ of real video, and with it $0.896$, above it (Appendix~\ref{sec:appendix_generation}). The frame index thus controls temporal variation, while the EEG controls content.

\begin{figure*}[!t]
\centering
\begin{minipage}[b]{0.245\linewidth}
  \centering
  \centerline{\includegraphics[width=\linewidth]{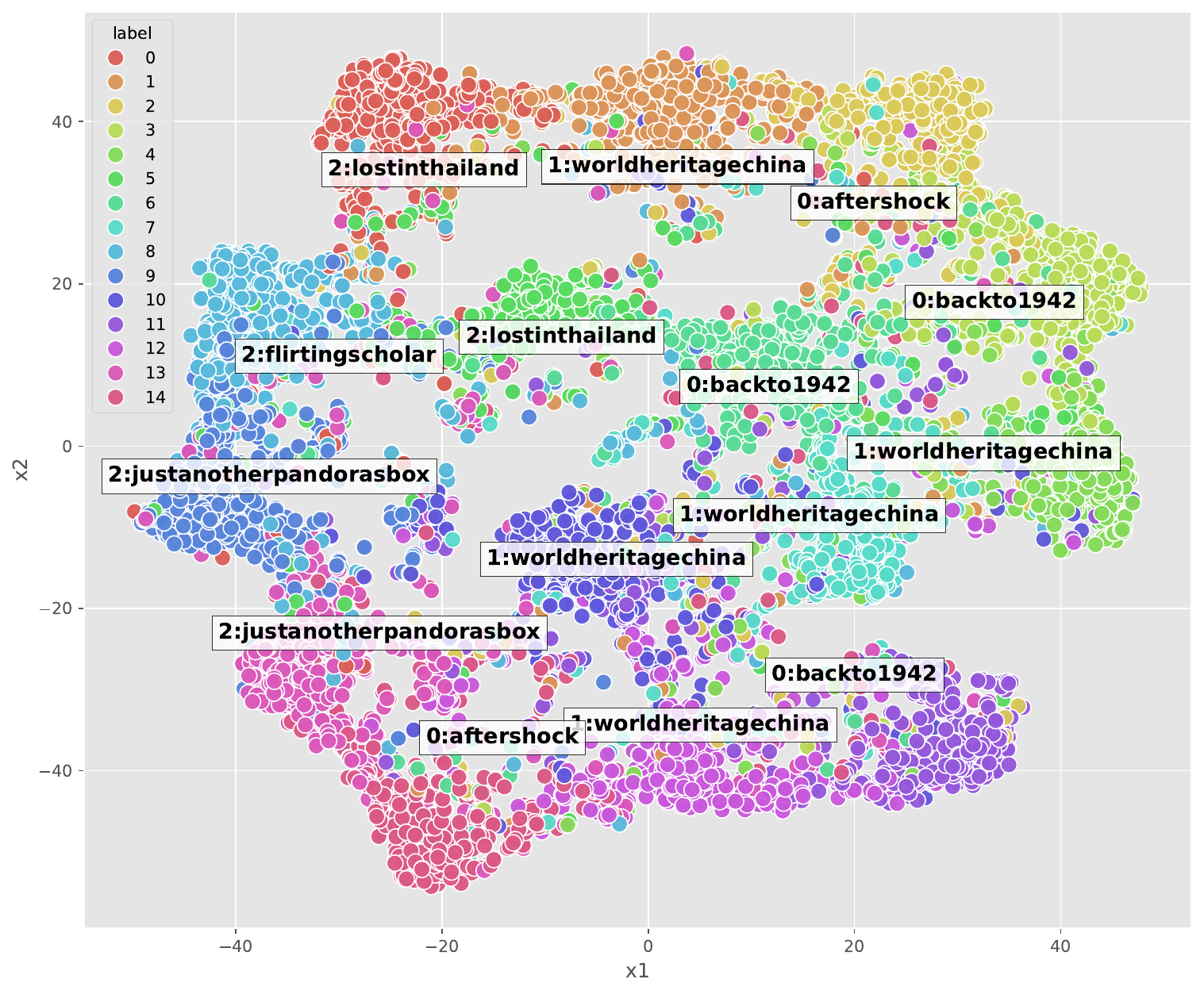}}\vspace{-0.2em}
  \centerline{(a)}\medskip
  \vspace{-4mm}
\end{minipage}
\hfill
\begin{minipage}[b]{.245\linewidth}
  \centering
  \centerline{\includegraphics[width=\linewidth]{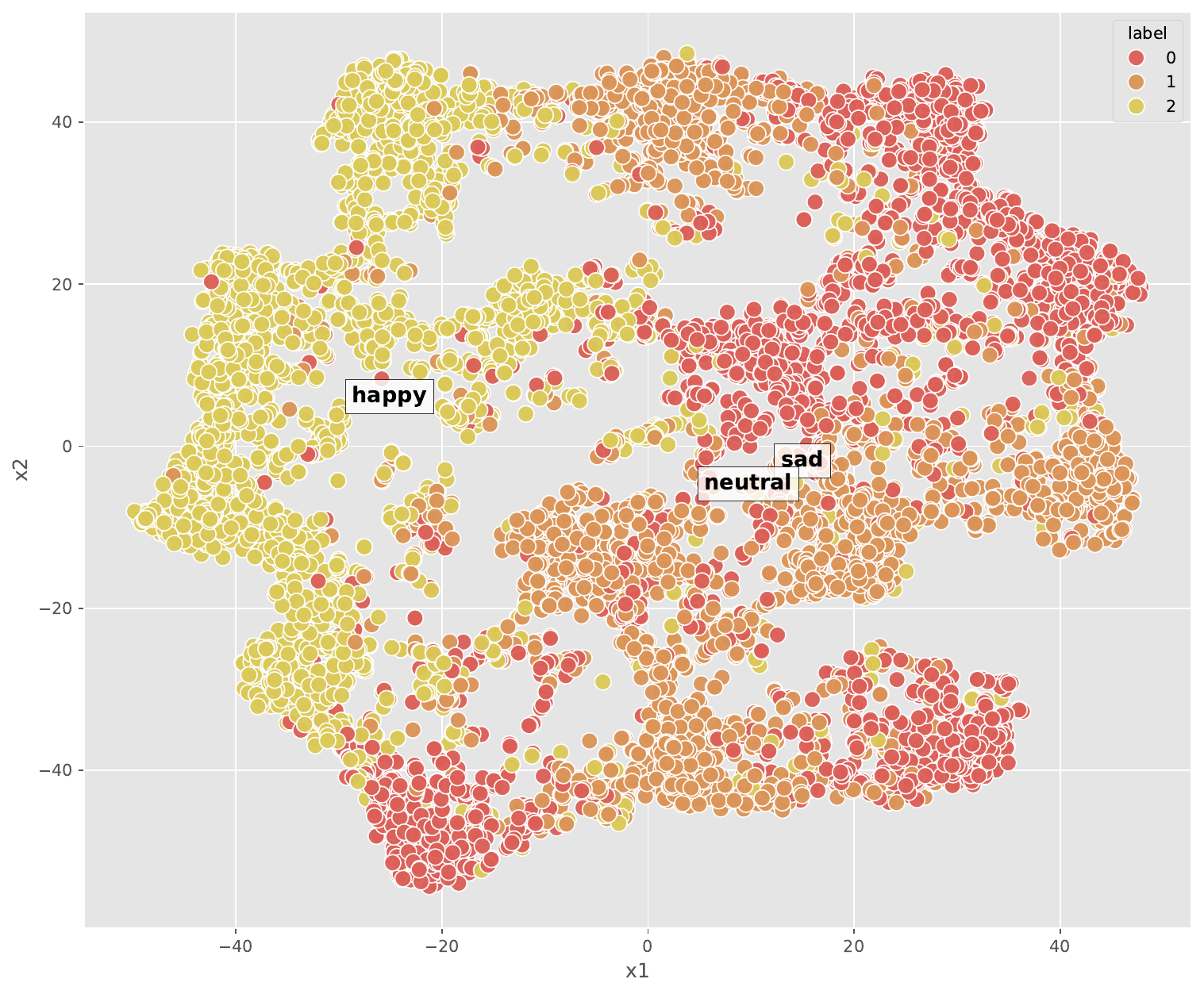}}\vspace{-0.2em}
  \centerline{(b)}\medskip
 \vspace{-4mm}
\end{minipage}
\hfill
\begin{minipage}[b]{0.245\linewidth}
  \centering
  \centerline{\includegraphics[width=\linewidth]{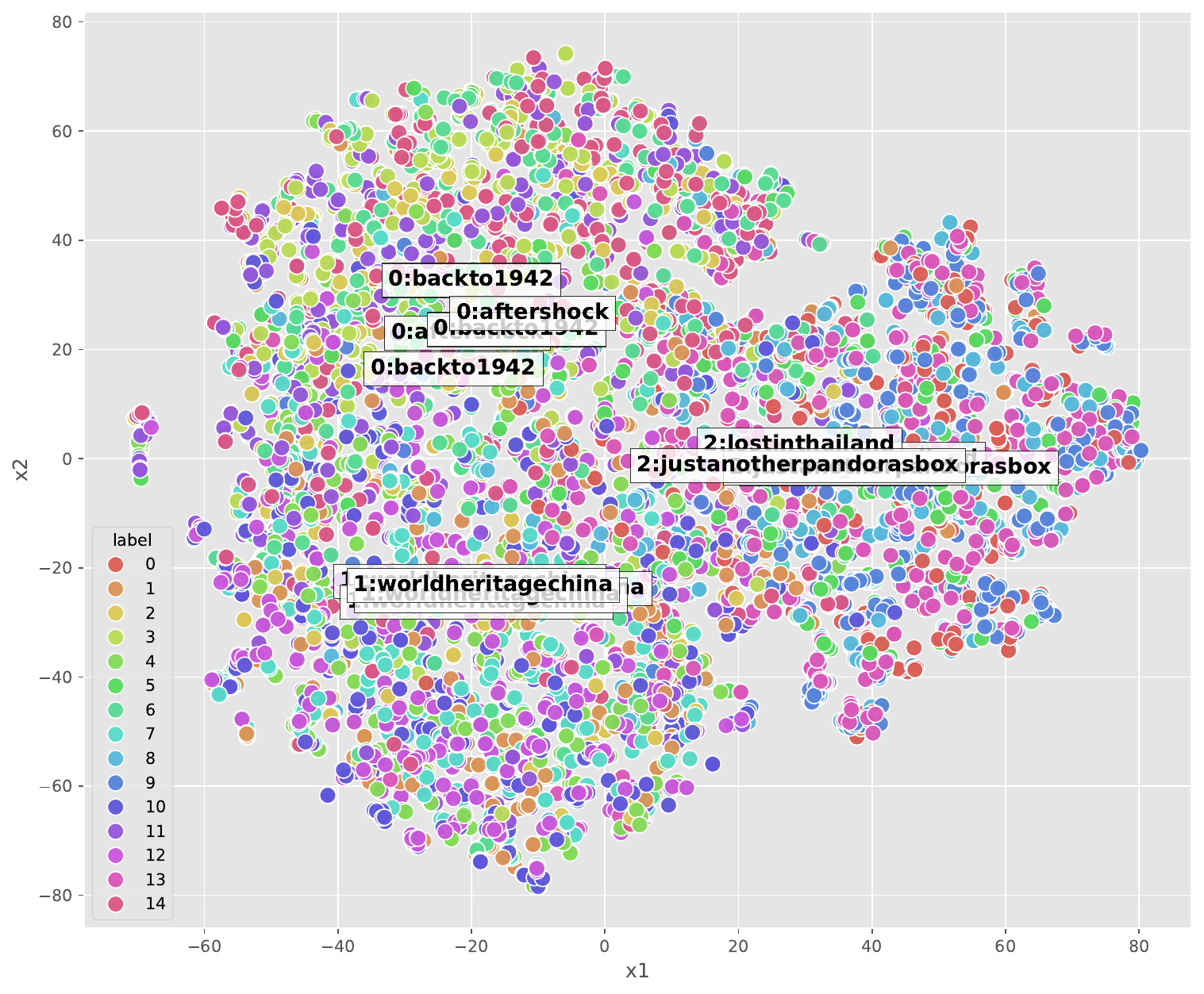}}\vspace{-0.2em}
  \centerline{(c)}\medskip
 \vspace{-4mm}
\end{minipage}
\hfill
\begin{minipage}[b]{0.245\linewidth}
  \centering
  \centerline{\includegraphics[width=\linewidth]{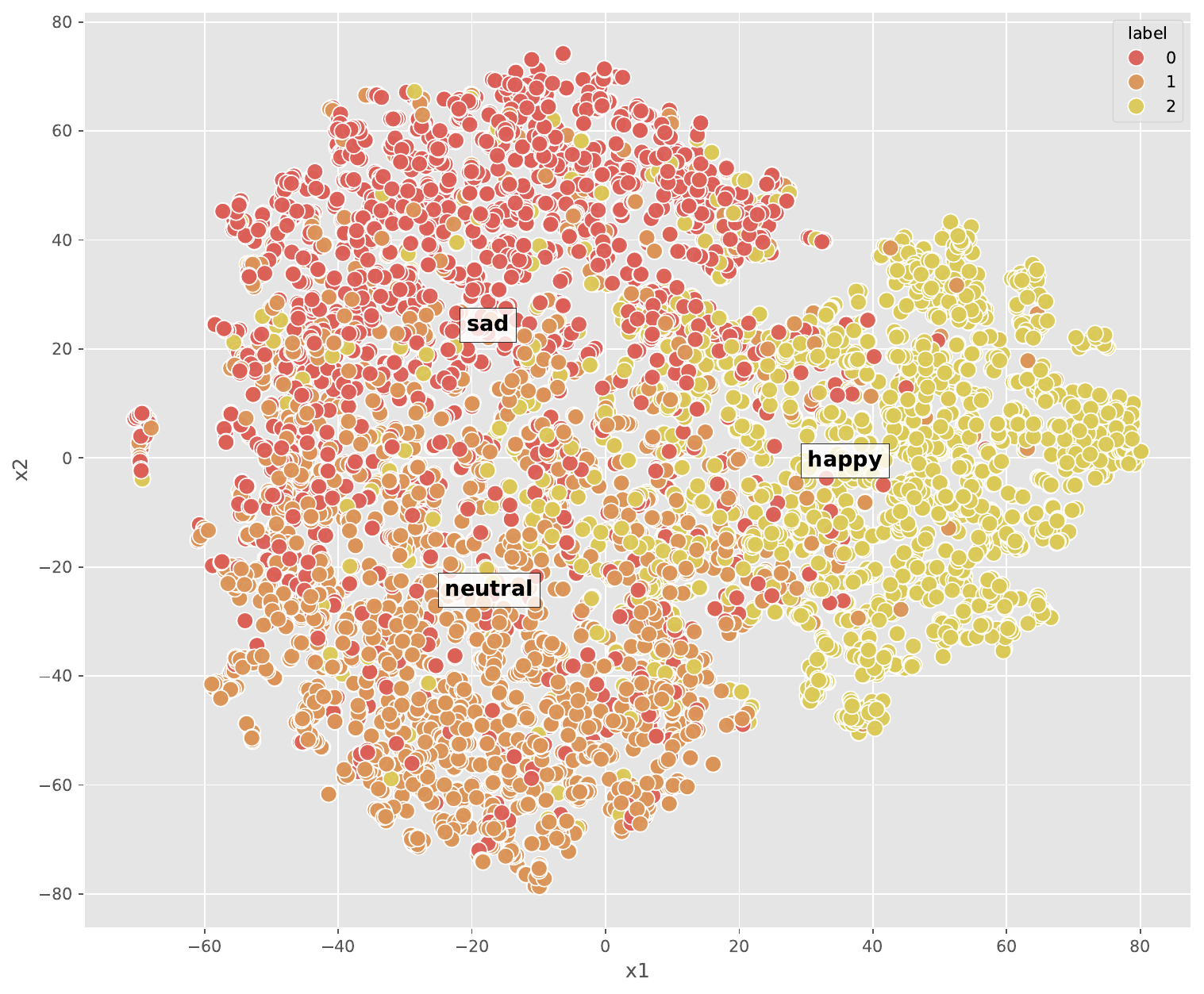}}\vspace{-0.2em}
  \centerline{(d)}\medskip
 \vspace{-4mm}
\end{minipage}\vspace{-0.5em}
\caption{\textbf{Interaction of visual and emotional features in EEG.} EEG embeddings from the SEED dataset under two training regimes. (a-b) Training with video labels: embeddings cluster by video stimulus and, when colored by emotion, also separate by emotional category, showing visual representations retain emotional information. (c-d) Training with emotion labels: embeddings cluster by emotional state but fail to discriminate between video stimuli. EEG thus multiplexes visual and emotional content, with the dominant factor set by the training objective, motivating our use of emotion classification as supporting evidence for dynamic video reconstruction. The plots label the positive and negative emotions as happy and sad. Zoom in for detail.}
\label{fig:tsne_plots}
\vspace{-1.2em}
\end{figure*}

EEGStyleGAN-ADA is trained only on dataset-specific videos from SEED \citep{duan2013differential, zheng2015investigating}, EEG-Video Action \citep{yao2024identifying}, and SEED-DV \citep{eeg2video}, without pre-training. The EEG feature provides semantic guidance, while the GAN provides the distributional prior needed for frame synthesis. Figure~\ref{fig:teaser}(a) and Appendix~\ref{sec:appendix_qual} show generated frames, and Table~\ref{table:quant_gan_eval} reports their quantitative quality. The segments used to train and evaluate this generator come from the 80/10/10 random split described in Section~\ref{sec:experiment}, for all three datasets. The training and test sets therefore contain different 2-second chunks of the same videos, so Table~\ref{table:quant_gan_eval} measures in-distribution reconstruction. Generalization to unseen stimuli is evaluated at the encoder level in Section~\ref{sec:diagnostics}.

\section{Experiments}
\label{sec:experiment}

In this section, we study a) the effect of the training labels on EEG feature learning with triplet loss, b) the information encoded by different brain regions and time windows, and c) whether the EEG-conditioned generator decodes segment-specific content, using controlled diagnostics. In all experiments, the input to the network is EEG with shape $\mathbb{R}^{c \times t}$, where $c$ is the number of channels and $t$ is the total number of timesteps. The network produces a unit-normalized feature $\psi \in \mathbb{R}^{1024}$, evaluated with $k$-means classification accuracy \citep{hartigan1979algorithm} or, on SEED-DV, linear probing. Details are in Appendix~\ref{sec:appendix_implement}.

\begin{table}[!t]
\centering
\small
\caption{\textbf{Quantitative evaluation of generated video frames.} In-distribution evaluation on the random test split of each dataset. CLIP-Sim is the cosine similarity between CLIP embeddings of generated and ground-truth frames. The EEG2Video SSIM is the value reported by \citet{eeg2video}.}\vspace{-0.5em}
\resizebox{0.9\linewidth}{!}{
\def\arraystretch{1.2}
\begin{tabular}{l|l|c|c|c|c}
\hline
Dataset & Model & PSNR $\uparrow$ & SSIM $\uparrow$ & LPIPS $\downarrow$ & CLIP-Sim $\uparrow$ \\ \hline
SEED \citep{duan2013differential, zheng2015investigating} & Ours & 10.80 & 0.153 & 0.587 & 0.679 $\pm$ 0.037 \\
EEG-Video Action \citep{yao2024identifying} & Ours & 16.22 & 0.431 & 0.418 & 0.772 $\pm$ 0.022 \\ \hline
\multirow{2}{*}{SEED-DV \citep{eeg2video}} & EEG2Video \citep{eeg2video} & -- & 0.256 & -- & -- \\
 & Ours & 9.62 & 0.143 & 0.690 & 0.558 $\pm$ 0.023 \\ \hline
\end{tabular}}
\label{table:quant_gan_eval}\vspace{-0.5em}
\end{table}

\begin{table}[!t]
\centering
\small
\caption{\textbf{Datasets.} All EEG is resampled to $200$\,Hz and cut into non-overlapping $2$-second segments. The controlled diagnostics (Section~\ref{sec:diagnostics}) use EEG-Video Action only.}\vspace{-0.5em}
\resizebox{0.9\linewidth}{!}{
\def\arraystretch{1.15}
\begin{tabular}{l|c|c|c|c}
\hline
Dataset & Subjects & Stimuli & Labels & Channels \\ \hline
SEED \citep{duan2013differential,zheng2015investigating} & 15 ($\times$3 sessions) & 15 clips, $\sim$4 min each & 15 videos / 3 emotions & 62 \\
EEG-Video Action \citep{yao2024identifying} & 19 & 13 single-shot videos & 13 videos & 64 \\
SEED-DV \citep{eeg2video} & 20 & 1400 clips & 40 concepts & 62 \\ \hline
\end{tabular}}
\label{tab:datasets}
\end{table}

\vspace{1mm} \noindent
\textbf{Datasets.} We use three EEG-video datasets (Table~\ref{tab:datasets}): SEED \citep{duan2013differential, zheng2015investigating}, EEG-Video Action \citep{yao2024identifying}, and SEED-DV \citep{eeg2video}. All EEG is band-limited and resampled to $200$\,Hz, then cut into non-overlapping $2$-second segments ($400$ timesteps) aligned with the corresponding video frames. Unless stated otherwise, the segments of each video are randomly split into $80\%$ training, $10\%$ validation, and $10\%$ test. Train and test therefore contain different segments of the same videos and subjects. Further details are in Appendix~\ref{sec:appendix_datasets}.

\subsection{What does an EEG encode?}
\label{subsec:what_encode}

We first study whether triplet loss \citep{schroff2015facenet} can learn meaningful representations from EEG signals. On the EEG-Video Action dataset \citep{yao2024identifying}, we extract EEG features using a randomly initialized network and cluster them by subject labels and video classes. As shown in Figure~\ref{fig:tsne_plots_single_shot}(a), the subject clusters are well separated before training, showing that EEG contains information about subject identity. In contrast, the video clusters in Figure~\ref{fig:tsne_plots_single_shot}(b) are not well separated, indicating that the untrained features contain less information about the observed videos. We then train the network using video labels and triplet loss. As shown in Figure~\ref{fig:tsne_plots_single_shot}(c), the learned features form distinct clusters for different video stimuli. When the same features are grouped by subject ID, the clusters collapse into a single cluster, as shown in Figure~\ref{fig:tsne_plots_single_shot}(d). This shows that triplet loss shifts the learned representation toward stimulus-related information while reducing subject-related structure. Section~\ref{sec:diagnostics} tests how far this reduction goes on unseen videos.

The SEED dataset contains both emotion and video class labels. We further analyze the learned representations in Figure~\ref{fig:tsne_plots}. In Figures~\ref{fig:tsne_plots}(a-b), the network is trained using video labels. The embeddings form clear clusters based on the video stimuli. When we color the same embeddings by emotion label, videos with the same emotion lie close together. Because each video carries a single emotion label, this is not implied by the video clusters alone. It indicates that EEG representations trained on video labels contain implicit emotional information. In Figures~\ref{fig:tsne_plots}(c-d), the network is instead trained using emotion labels. The embeddings then cluster primarily by emotion and fail to clearly distinguish among different video stimuli. This shows that emotion-based supervision does not preserve the same fine-grained visual information as video-based supervision.

\subsection{How do brain regions affect EEG decoding?}
\label{subsec:brain_region}

We study EEG classification across different brain regions for two tasks. The first is video classification on the SEED \citep{duan2013differential, zheng2015investigating}, EEG-Video Action \citep{yao2024identifying}, and SEED-DV \citep{eeg2video} datasets. The second is emotion classification with $3$ classes on SEED. For each experiment, the EEG feature extraction network is trained using only channels from the selected brain region. We evaluate two regimes. $R1$) All Subjects, where training, validation, and testing use non-overlapping segments from the same subjects. $R2$) Leave Two Subjects, where the model is trained on $S-2$ subjects and tested on the two remaining subjects, $(S-1)$ and $S$, where $S$ is the total number of subjects.

\begin{figure*}[tp]
    \centering
    \includegraphics[width=0.95\linewidth]{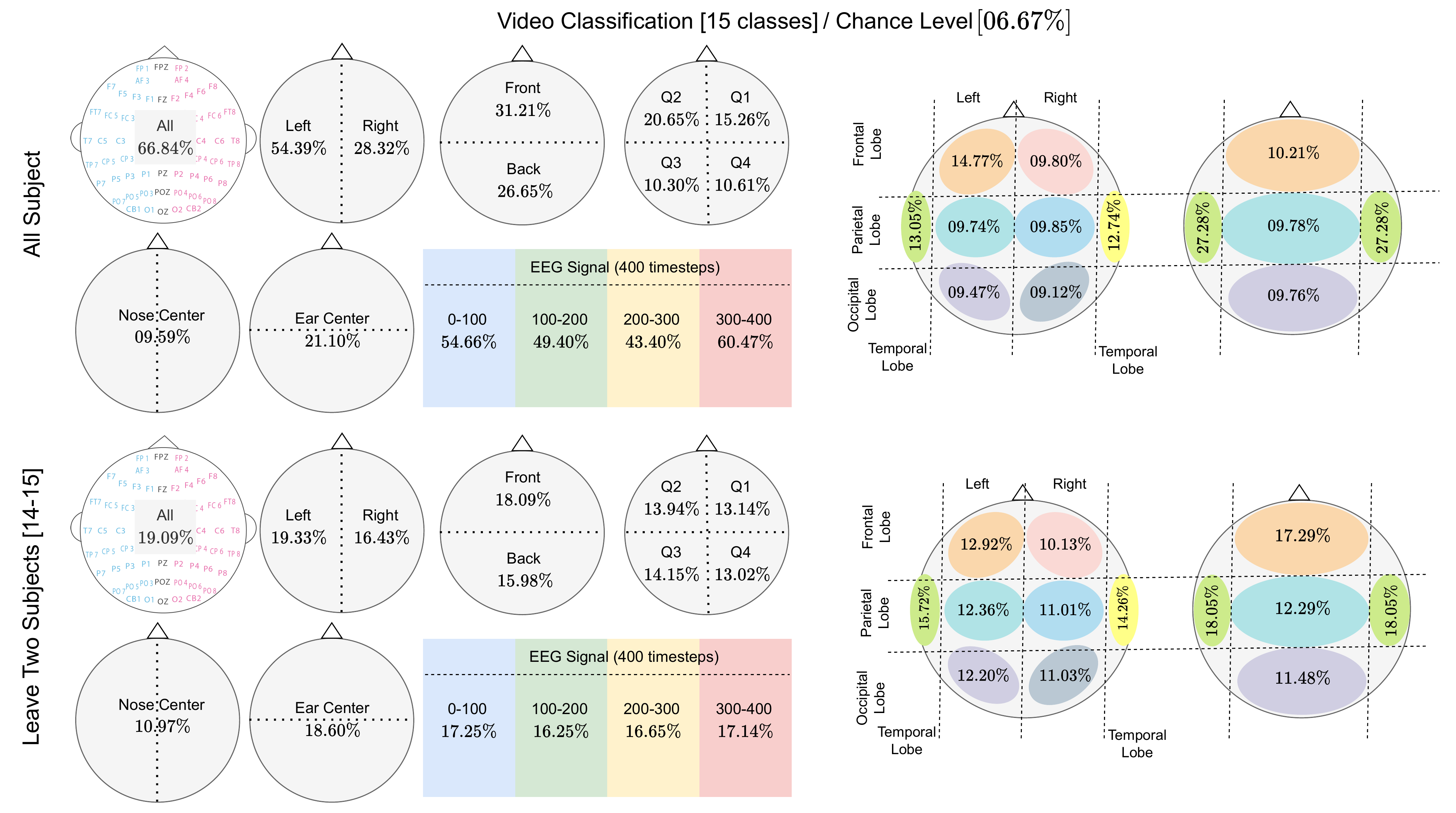}\vspace{-0.8em}
    \caption{\textbf{Video Decoding on SEED}. The figure shows the test classification accuracy of EEG signals when trained on different brain region channels for video classification on the SEED dataset \citep{duan2013differential, zheng2015investigating}. The two rows show the two evaluation regimes. Under All Subjects, training, validation and test segments come from the same recordings but do not overlap. Under Leave Two Subjects, the network is trained on the data of subjects $1$ to $13$ and tested on subjects $14$ and $15$ from the SEED dataset. To study the effect of the time segment, we use a pre-trained network, mask one $100$-timestep window, and compute the accuracy with all channels.}
    \label{fig:within_leavetwo_video_emotion}\vspace{-1.0em}
\end{figure*}

\vspace{1mm} \noindent
\textbf{\textit{Video Classification on SEED Dataset}}. We first classify the $15$ video stimuli using all EEG channels. Under the All Subjects regime, the model achieves an overall accuracy of $66.84\%$, well above the chance level of $6.67\%$. This shows that EEG contains meaningful information about the observed video stimuli, as shown in the first row of Figure~\ref{fig:within_leavetwo_video_emotion}. We then divide the EEG channels into brain regions to study their contribution. The Left Hemisphere achieves higher accuracy than the Right Hemisphere ($54.39\%$ vs.\ $28.32\%$). The Front Region also outperforms the Back Region ($31.21\%$ vs.\ $26.65\%$).

Under the Leave Two Subjects regime, the overall accuracy drops to $19.09\%$, reflecting the difficulty of cross-subject generalization due to differences in EEG patterns. However, the accuracy remains nearly $3\times$ above chance. The Left Hemisphere continues to outperform the Right Hemisphere ($19.33\%$ vs.\ $16.43\%$). These results show that the learned representation retains some stimulus-related information across subjects, although its accuracy is reduced. A split by time instead of by subject is reported in Appendix~\ref{sec:appendix_time}.

\begin{figure*}[!t]
    \centering
    \includegraphics[width=0.95\linewidth]{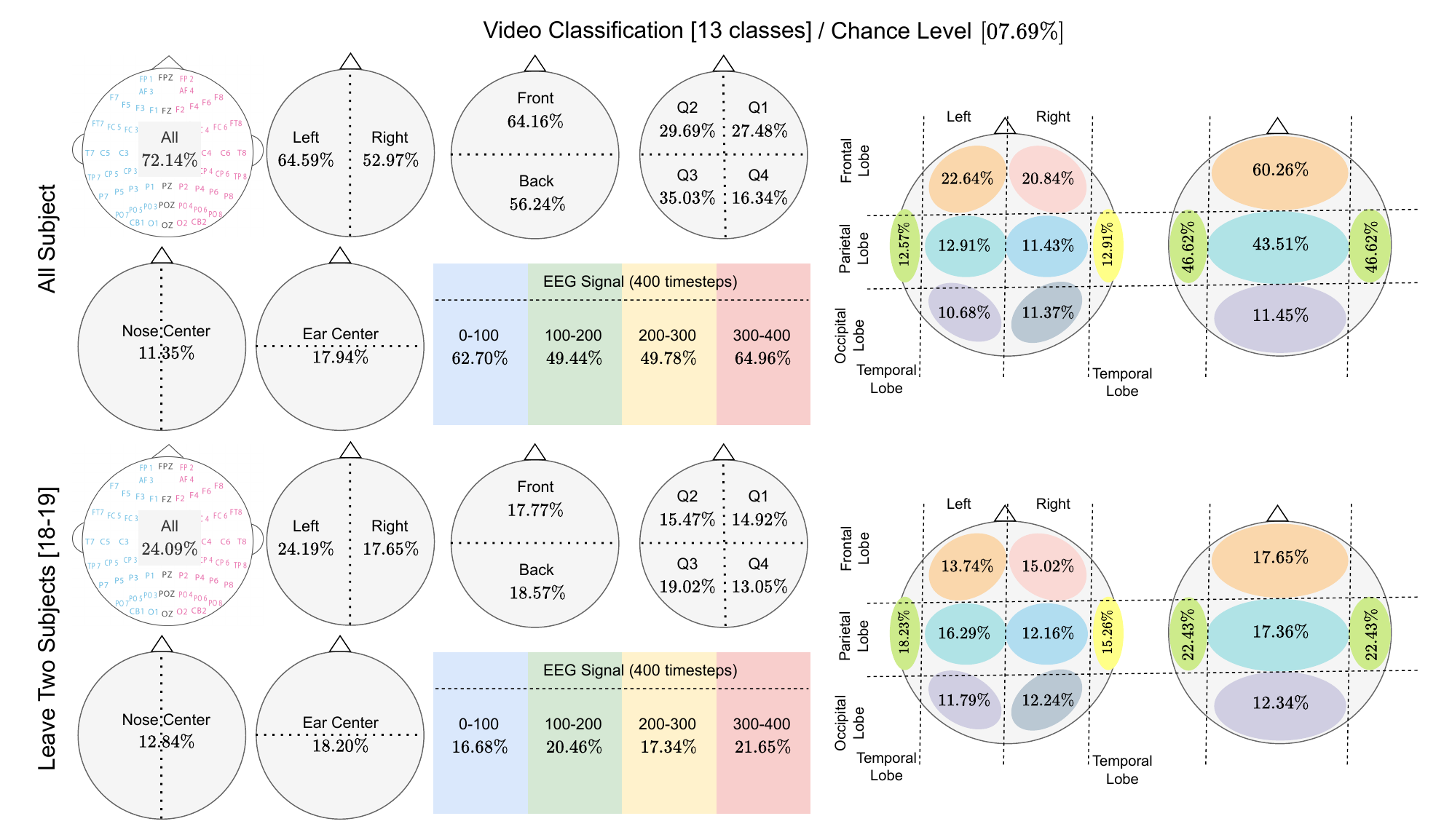}\vspace{-1.0em}
    \caption{\textbf{Video Decoding on EEG-Video Action.} The two rows show the All Subjects and Leave Two Subjects (subjects $18$ and $19$ held out) regimes. The figure illustrates the effect of sampling EEG channels from specific brain regions and performing training and testing of the EEG feature extraction network over the video labels from the EEG-Video Action dataset \citep{yao2024identifying}. It also shows the effect of different non-overlapping time segments in the pre-trained networks across all channels, i.e., how classification changes when a particular time region ($t_{1}$-$t_{2}$) is masked.} \vspace{-1.0em}
    \label{fig:singleshot_collage}
\end{figure*}

\vspace{1mm} \noindent
\textbf{\textit{Video Classification on EEG-Video Action Dataset}}. Figure~\ref{fig:singleshot_collage} shows the video classification results on the EEG-Video Action dataset \citep{yao2024identifying}. Under the All Subjects regime, the model achieves $72.14\%$ accuracy across $13$ classes, well above the chance level of $7.69\%$. The Left Hemisphere outperforms the Right Hemisphere ($64.59\%$ vs.\ $52.97\%$), consistent with the hemispheric asymmetry observed on the SEED dataset. The Front Region also outperforms the Back Region ($64.16\%$ vs.\ $56.24\%$), showing a similar regional pattern across the two datasets. Under the Leave Two Subjects regime, accuracy drops to $24.09\%$, again showing the difficulty of cross-subject generalization. The Left Hemisphere continues to outperform the Right Hemisphere ($24.19\%$ vs.\ $17.65\%$).

\vspace{1mm} \noindent
\textbf{\textit{Emotion Classification on SEED Dataset}}. For three-way emotion classification (positive, neutral, negative), the model reaches $76.33\%$ under the All Subjects regime (chance $33.33\%$) and $54.80\%$ under Leave Two Subjects (Appendix~\ref{sec:appendix_seedemotion}). The hemispheric pattern reverses relative to video classification. The Right Hemisphere outperforms the Left ($68.16\%$ vs.\ $49.54\%$), and this holds for unseen subjects ($57.07\%$ vs.\ $52.86\%$). Under the All Subjects regime, the Front Region again outperforms the Back Region ($67.75\%$ vs.\ $62.36\%$). For unseen subjects, this front-back difference does not hold for emotion ($48.79\%$ vs.\ $50.79\%$) or for EEG-Video Action video classification ($17.77\%$ vs.\ $18.57\%$), although it does for SEED video classification ($18.09\%$ vs.\ $15.98\%$).

\vspace{1mm} \noindent
\textbf{\textit{Video Classification on SEED-DV Dataset}}. We perform brain region analysis on one subject (sub1) from SEED-DV using the $7$-fold protocol of \citet{eeg2video}. All accuracies are near chance ($2.48\%$ to $4.71\%$, chance $2.5\%$), and masking some windows even raises accuracy above the unmasked $3.71\%$, so these results do not show regional differences on SEED-DV (Figure~\ref{fig:seed_dv_topomap}, Appendix~\ref{sec:appendix_seedandseeddv}).

\vspace{1mm} \noindent
\textbf{\textit{Effect of Different Brain Lobes.}} We further train and evaluate the model using only channels from each brain lobe. The highlighted regions in Figures~\ref{fig:within_leavetwo_video_emotion} and~\ref{fig:singleshot_collage} are approximations of the brain region (Appendix~\ref{sec:appendix_section_eeg_montage_cap}). With only six channels, the temporal lobe gives the highest accuracy in five of the six settings: SEED video classification ($27.28\%$ against at most $10.21\%$ for other lobes, and $18.05\%$ for unseen subjects, $0.76$ points above the frontal lobe), SEED emotion classification ($70.85\%$ and $60.43\%$), and EEG-Video Action for unseen subjects ($22.43\%$). For EEG-Video Action with overlapping subjects, the $24$-channel frontal lobe leads ($60.26\%$), followed by the temporal lobe ($46.62\%$).

\vspace{1mm} \noindent
\textbf{\textit{Temporal Dynamics.}} Figures~\ref{fig:within_leavetwo_video_emotion} and~\ref{fig:singleshot_collage} show how classification changes across EEG timesteps for the video classification task. The network is first trained using video labels. During inference, we mask the EEG timesteps between $t_{1}$ and $t_{2}$ and extract features from the masked signal. We then perform $k$-means \citep{hartigan1979algorithm} classification on these features. Under the All Subjects regime, classification accuracy drops most on both datasets when timesteps $100$--$300$ are masked, which corresponds to roughly $500$--$1500$\,ms within each $2$-second window (SEED: $54.66$, $49.40$, $43.40$, $60.47\%$ for the four masked windows; EEG-Video Action: $62.70$, $49.44$, $49.78$, $64.96\%$). This interval carries the most stimulus information for the classifier. Under Leave Two Subjects, the effect is smaller on SEED and not observed on EEG-Video Action.

The lower performance under the Leave Two Subjects protocol reflects the known inter-subject variability of EEG signals. Although domain adaptation could reduce this gap, our goal is to study the information encoded in the EEG under controlled conditions. The hemispheric pattern holds in both regimes, whereas the front-back effects and, for video classification, the drop when timesteps $100$--$300$ are masked largely disappear for unseen subjects.

\subsection{Controlled Diagnostics for EEG-Conditioned Video Generation} \label{sec:diagnostics}

Reconstruction quality on a fixed training distribution cannot, by itself, distinguish segment-specific EEG decoding from coarse category information combined with a generator that has memorized a small set of training videos. We therefore run one sanity check and three controlled experiments on the EEG-Video Action dataset~\citep{yao2024identifying}.

\vspace{1mm} \noindent
\textbf{Terminology.} A \textit{trial} is one distinct video stimulus, shown once to every subject. The dataset has thirteen trials, so we use \textit{trial}, \textit{stimulus identity}, and \textit{class} interchangeably. \textit{Segment-specific} decoding means recovering what a particular segment shows, beyond the identity of its trial. A \textit{segment} is one $2$-second EEG window with its video frames. In a mismatched-conditioning test, the true trial of a segment is the \textit{original} and the substituted trial is the \textit{donor}. A \textit{class prototype} is the mean CLIP \citep{radford2021learning} image embedding of all ground-truth test frames of one class, $\ell_2$-normalized. The \textit{nearest-prototype vote} (NPV) assigns each generated frame to the class whose prototype has the highest cosine similarity to it. NPV-D and NPV-O are the fractions of frames assigned to the donor and original class.

\begin{figure}[!t]
    \centering
    \includegraphics[width=0.7\linewidth]{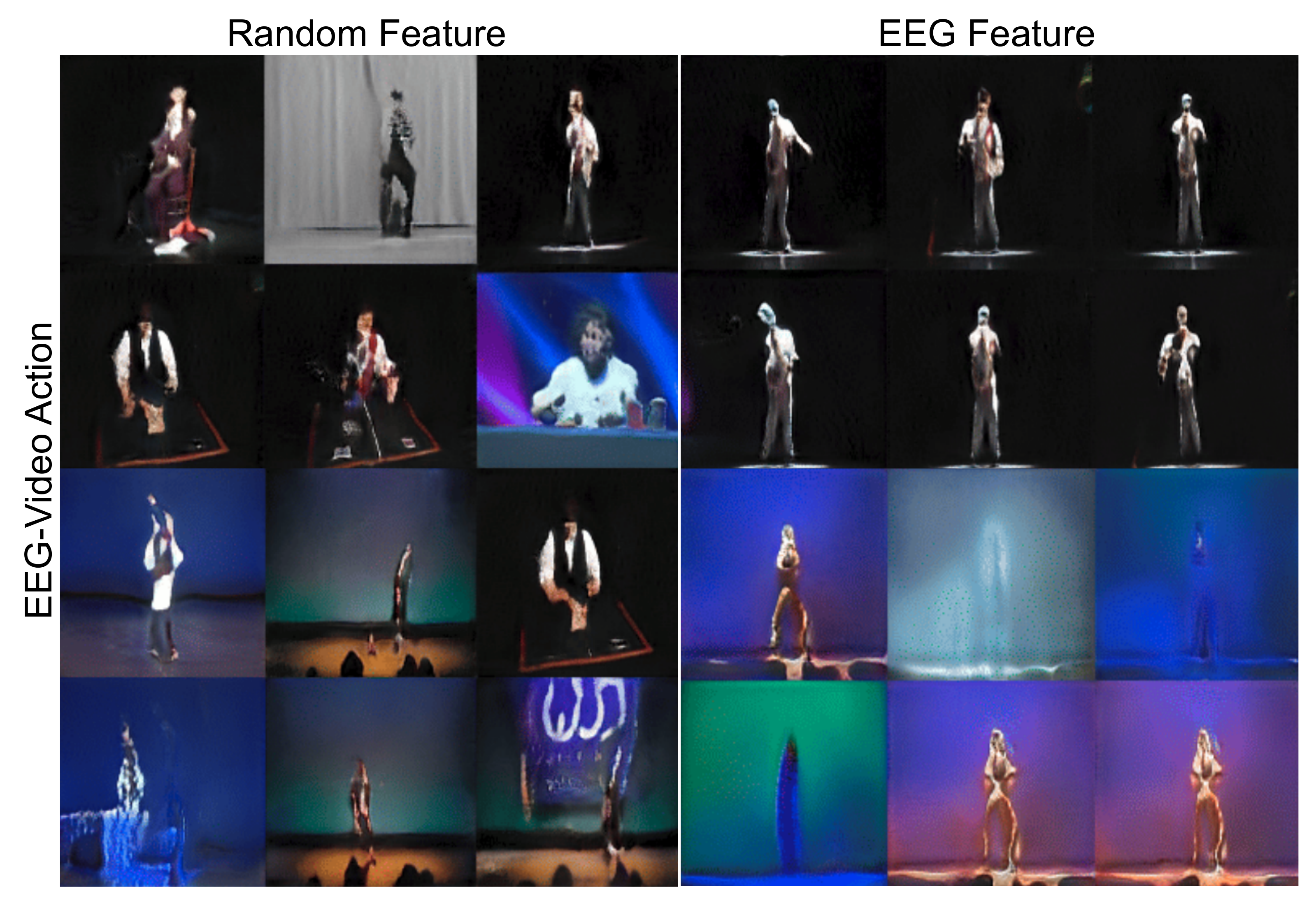}
    \caption{\textbf{Random conditioning control} (EEG-Video Action). Frames generated from random vectors (left) and from learned EEG embeddings (right). Random conditioning gives frames without class-consistent content, while EEG conditioning gives consistent, stimulus-aligned content. This shows the generator needs structured EEG input. Which trial's EEG it follows is tested separately in Table~\ref{tab:exp3mismatch}.}
    \label{fig:randomfeat_eegfeat}\vspace{-1.0em}
\end{figure}

\vspace{1mm} \noindent
\textbf{Sanity check (random conditioning).} We first replace the learned EEG embeddings with random vectors sampled from $\mathcal{N}(0, I)$, keeping the generator and all other settings unchanged. Random conditioning fails to produce class-consistent outputs, whereas real EEG embeddings yield consistent content across frames (Figure~\ref{fig:randomfeat_eegfeat}). This shows that the generator relies on structured EEG input. The mismatched-EEG control below further tests whether it follows the specific trial encoded in the EEG.

\vspace{1mm} \noindent
\textbf{Held-out-stimulus generalization of the EEG encoder.} To test whether the EEG encoder captures video identity beyond the classes used for training, we retrained the triplet-loss encoder four times, each time excluding a different set of trial classes from training and validation. We then embedded all EEG segments from the excluded classes and compared them with features from a randomly initialized encoder of the same architecture. We evaluated two properties. First, we measured whether held-out classes formed clusters that matched their true video identities, using $k$-means accuracy averaged over $20$ random seeds. Second, we measured whether the nearest neighbor of each held-out segment, based on cosine similarity, belonged to the same video. We report this both across all pairs and for pairs from different subjects only. The cross-subject result is more informative because EEG segments from the same subject share physiological patterns that are unrelated to the video content. These patterns can inflate retrieval accuracy by allowing the encoder to recognize subjects rather than videos. Table~\ref{tab:exp1} reports results for four non-overlapping held-out splits covering all thirteen trial classes.

The $k$-means results are consistent across all four splits. The trained encoder has a higher mean accuracy than the random encoder on every split, by $1.3$ to $6.7$ points, although the margin is small on split 9, 10, 11, 12. This provides a consistent signal that the encoder captures video-related structure in unseen classes. All-pairs retrieval shows a large gap between trained and random encoders ($16$--$21$ points). Restricting matches to different subjects separates stimulus information from subject identity: $65$--$72\%$ of correct all-pairs matches come from the same subject as the query, and the cross-subject gap is $1.4$ points on average (from $0.1$ points on split 0, 1, 2 up to $3.1$ points on split 6, 7, 8). This mirrors the cross-subject gap of the Leave Two Subjects study and identifies cross-subject alignment as the main lever for improving held-out generalization.

\vspace{1mm} \noindent
\textbf{Mismatched-EEG control.} For each test segment, we replace its EEG conditioning with EEG from a different, randomly selected donor trial at the same frame position, keeping the positional encoding unchanged. We then measure the change in reconstruction quality and whether the generated content shifts toward the donor. Reconstruction quality drops sharply under mismatched conditioning (Table~\ref{tab:exp3mismatch}), showing that the generator does not ignore the EEG. More directly, the mean CLIP similarity of mismatched outputs to the donor prototype ($0.893$) exceeds their similarity to the original prototype ($0.754$) by $0.139$. The nearest-prototype vote assigns $82.3\%$ of mismatched outputs to the donor class, $1.4\%$ to the original class, and $16.4\%$ to neither. The generator therefore uses EEG as a content selector rather than producing a fixed, EEG-independent output, and follows the stimulus class carried by the EEG.

\vspace{1mm} \noindent
\textbf{Class-conditional baseline.} To isolate the operational value of EEG conditioning from the generator's dataset-specific prior, we train an otherwise identical generator using a one-hot ground-truth trial label instead of the EEG feature, keeping the frame-position encoding. We then compare it with the EEG-conditioned model on the same test split. The class-conditional model outperforms the EEG-conditioned model on every metric (Table~\ref{tab:exp3main}). Two asymmetries between the models shape this comparison.

The first is a \emph{training-target asymmetry}. A one-hot label is a discrete, noise-free thirteen-way signal, whereas the 1024-dimensional EEG embedding is continuous and noisy. We repeat the mismatched-conditioning experiment on the class-conditional model by replacing the correct label with an incorrect one. Reconstruction falls to a similar level (PSNR $10.12$ against $10.34$), a larger drop from its higher start, and the class-conditional model follows the swapped label more strongly, with $94.3\%$ of outputs assigned to the donor class against $82.3\%$ for EEG (Table~\ref{tab:exp3mismatch}). The cleaner label is easier for the generator to bind to, independent of how much information either signal carries.

\begin{table}[!t]
\centering
\caption{\textbf{Held-out encoder generalization.} Results across four non-overlapping held-out splits that together cover all thirteen trial classes. The gap between trained and random columns, rather than either value alone, indicates generalization. $K$-means accuracy is averaged over twenty seeds, chance is $33.3\%$ for the three-class splits and $25.0\%$ for the four-class split. Cross-subject retrieval restricts nearest-neighbor matching to pairs from different subjects. Same-subject share is the fraction of correct all-pairs matches that come from the same subject as the query.}\vspace{-0.5em}
\resizebox{\linewidth}{!}{
\def\arraystretch{1.2}
\begin{tabular}{c|cc|cc|cc|c}
\hline
\multirow{2}{*}{Held-out split} & \multicolumn{2}{c|}{K-means acc.} & \multicolumn{2}{c|}{All-pairs retrieval} & \multicolumn{2}{c|}{Cross-subj.\ retrieval} & \multirow{2}{*}{Same-subj.\ share} \\
 & Trained & Random & Trained & Random & Trained & Random & \\ \hline
0, 1, 2       & 41.5\% & 35.9\% & 58.4\% & 37.1\% & 35.7\% & 35.6\% & 72.1\% \\
3, 4, 5       & 39.0\% & 36.5\% & 53.9\% & 36.1\% & 35.0\% & 33.2\% & 71.0\% \\
6, 7, 8       & 42.5\% & 35.8\% & 54.7\% & 38.6\% & 38.7\% & 35.6\% & 65.3\% \\
9, 10, 11, 12 & 28.7\% & 27.4\% & 43.8\% & 27.7\% & 26.7\% & 26.0\% & 67.5\% \\ \hline
\end{tabular}}\vspace{-0.5em}
\label{tab:exp1}
\end{table}


\begin{table*}[!t]
\centering
\begin{minipage}[t]{0.49\textwidth}
\centering
\caption{\textbf{Mismatched-conditioning control.} EEG-swap versus label-swap under mismatched conditioning, with correctly conditioned EEG as reference. NPV-D and NPV-O are the \textit{nearest-prototype vote} rates for the donor and original classes.}\vspace{-0.5em}
\resizebox{\linewidth}{!}{
\def\arraystretch{1.4}
\begin{tabular}{c|c|c|c|c|c}
\hline
Model & PSNR $\uparrow$ & SSIM $\uparrow$ & LPIPS $\downarrow$ & NPV-D $\uparrow$ & NPV-O $\downarrow$ \\ \hline
EEG (correct) & 16.22 & 0.431 & 0.418 & -- & -- \\ \hline
EEG-swap & 10.34 & 0.203 & 0.584 & 82.3\% & 1.4\% \\
Label-swap & 10.12 & 0.190 & 0.594 & 94.3\% & 0.4\% \\ \hline
\end{tabular}
}\vspace{-0.5em}
\label{tab:exp3mismatch}
\end{minipage}\hfill
\begin{minipage}[t]{0.49\textwidth}
\centering
\caption{\textbf{Class-conditional baseline vs.\ EEG-conditioned model.} Both models are evaluated on the same test split ($n = 30{,}096$ segments) using ground-truth class labels for the baseline.}\vspace{-0.5em}
\resizebox{\linewidth}{!}{
\def\arraystretch{1.4}
\begin{tabular}{c|c|c|c|c}
\hline
Model & PSNR $\uparrow$ & SSIM $\uparrow$ & LPIPS $\downarrow$ & CLIP-Sim $\uparrow$ \\ \hline
EEG-Cond. & 16.22 & 0.431 & 0.418 & 0.772 \\
Class-Cond. (G.T. label) & 17.13 & 0.620 & 0.285 & 0.866 \\ \hline
\end{tabular}
}\vspace{-0.5em}
\label{tab:exp3main}
\end{minipage}
\end{table*}

The second is a \emph{task asymmetry}. The label conditioning is identical for every segment of a trial, although one trial spans many visually distinct segments. The label model therefore cannot know which segment is wanted, and can only learn to sample plausible content for a class and frame position. To check that this does not reduce to collapse onto one fixed output, we generated eight frames per class-and-position cell from independent noise draws and compared their mean pairwise LPIPS with that of the real segments sharing the cell, over twenty cells. Generated diversity ($0.258$) nearly matches true diversity ($0.265$), a ratio of $0.97$ (Appendix~\ref{sec:appendix_classcond}). The label model thus samples genuine variation rather than repeating an archetype. The label therefore serves as a class-level reference for EEG conditioning, and matching it requires cleaner class-level EEG features, while going beyond it requires encoders that recover segment-level information from EEG.

\vspace{1mm} \noindent
\textbf{Summary.} Across the three experiments, EEG conditioning carries stimulus-level information that the generator actively uses as a content selector. The encoder separates unseen videos better than a randomly initialized encoder, the generator follows the class carried by the supplied EEG, and the ground-truth label provides a class-level reference point. Together, these experiments give a more precise account of the pipeline than any single reconstruction metric. Further analyses and ablations are in the appendix.

\section{Discussion}
\label{sec:discussion}

By learning EEG representations using a triplet objective, we reduced subject-specific structure on the training classes while preserving stimulus-related structure, although on unseen videos most correct matches remain within-subject. Classification accuracy was higher for left-hemisphere channels on video content, right-hemisphere channels on emotional content, and temporal-lobe channels in five of six SEED and EEG-Video Action settings. Under the All Subjects regime, accuracy also dropped when roughly $500$ to $1500$\,ms of each $2$-second window was masked at inference, making this the most informative interval for the classifier.

The controlled diagnostics of Section~\ref{sec:diagnostics} go beyond in-distribution reconstruction and show what the generator takes from the EEG. The EEG encoder separates entirely unseen videos better than a randomly initialized encoder across four held-out splits, although the gain across subjects is small ($1.4$ points), and substituting EEG from a different trial reliably shifts the generated output toward the donor trial's class. A class-conditional baseline that receives the ground-truth label scores higher on reconstruction metrics. The label is an easier training target, and it serves as a class-level reference that EEG encoders can be measured against.

Taken together, these results show that EEG carries stimulus-related information that the generator actively uses as a content selector, at the level of stimulus identity. We view this combination of controls, rather than reconstruction quality alone, as a more informative basis for evaluating EEG-conditioned video generation. The controls require no change to the generator and apply directly to the more elaborate architectures discussed in Section~\ref{sec:relatedwork}.

\vspace{1mm} \noindent
\textbf{Future directions.} Natural extensions of the protocol include datasets with repeated stimulus presentations, donor swaps within the same class, and generation for held-out videos. Encoders with stronger cross-subject alignment are the most direct route to finer-grained decoding.

\section{Conclusion}

We presented EEGVid to study what information about dynamic visual stimuli can be learned from EEG and how this information can support video generation. Visual and emotional information are jointly represented in EEG, with video-supervised representations retaining emotional structure while emotion-based supervision does not preserve the same visual information. Triplet learning also shifts EEG representations away from subject-specific structure toward stimulus-related information. Analysis across brain regions, hemispheres, and time further reveals consistent differences in visual and emotional encoding, with temporal regions contributing strongly across tasks. We then use these representations to generate short video clips with $8$ frames at $128 \times 128$ resolution using a modified StyleGAN-ADA conditioned on EEG features and temporal position embeddings. Three controlled diagnostics show that the EEG encoder separates unseen video classes better than a randomly initialized encoder, with a small cross-subject gain, and that replacing the EEG conditioning shifts generated content toward the substituted stimulus. Thus, the generator actively uses stimulus-related information in the EEG. Although a ground-truth class label yields stronger reconstruction metrics, follow-up experiments indicate that it also provides a cleaner, easier conditioning target. Overall, our results show that EEG contains a consistent visual and emotional structure that can support dynamic video generation, whereas the current pipeline primarily provides coarse stimulus-level information rather than fine-grained, segment-specific decoding.

\section*{Acknowledgement}
This work is supported by the Prime Minister Research Fellowship (PMRF2122-2557) to Prajwal Singh and the Dr.~Vilas Mujumdar Chair held by Shanmuganathan Raman. We also thank Muskan Priyadarshani at IIT Gandhinagar for her valuable input.

\bibliography{main}

@String(CVPR= {IEEE Conf. Comput. Vis. Pattern Recog.})

@String(ICCV= {Int. Conf. Comput. Vis.})

@String(ECCV= {Eur. Conf. Comput. Vis.})

@String(ICASSP=	{ICASSP})

@String(ICLR = {Int. Conf. Learn. Represent.})

@String(AAAI = {AAAI})

@String(CVPR  = {CVPR})

@String(ICCV  = {ICCV})

@String(ECCV  = {ECCV})

@String(ICLR  = {ICLR})

@article{wandell1999computational,
  title={Computational neuroimaging of human visual cortex},
  author={Wandell, Brian A},
  journal={Annual Review of Neuroscience},
  volume={22},
  number={1},
  pages={145--173},
  year={1999},
}

@book{fish2021philosophy,
  title={Philosophy of perception: A contemporary introduction},
  author={Fish, William},
  year={2021},
  publisher={Routledge}
}

@incollection{kosslyn1991cognitive,
  title={A cognitive neuroscience of visual cognition: Further developments},
  author={Kosslyn, Stephen M},
  booktitle={Advances in Psychology},
  volume={80},
  pages={351--381},
  year={1991},
  publisher={Elsevier}
}

@article{34,
  title={Visual image reconstruction from human brain activity using a combination of multiscale local image decoders},
  author={Miyawaki, Yoichi and Uchida, Hajime and Yamashita, Okito and Sato, Masa-aki and Morito, Yusuke and Tanabe, Hiroki C and Sadato, Norihiro and Kamitani, Yukiyasu},
  journal={Neuron},
  volume={60},
  number={5},
  pages={915--929},
  year={2008},
}

@article{35,
  title={Bayesian reconstruction of natural images from human brain activity},
  author={Naselaris, Thomas and Prenger, Ryan J and Kay, Kendrick N and Oliver, Michael and Gallant, Jack L},
  journal={Neuron},
  volume={63},
  number={6},
  pages={902--915},
  year={2009},
}

@article{36,
  title={Decoding and reconstructing color from responses in human visual cortex},
  author={Brouwer, Gijs Joost and Heeger, David J},
  journal={Journal of Neuroscience},
  volume={29},
  number={44},
  pages={13992--14003},
  year={2009},
}

@article{37,
  title={Linear reconstruction of perceived images from human brain activity},
  author={Schoenmakers, Sanne and Barth, Markus and Heskes, Tom and van Gerven, Marcel},
  journal={NeuroImage},
  volume={83},
  pages={951--961},
  year={2013},
}

@inproceedings{38,
  booktitle={Advances in Neural Information Processing Systems (NeurIPS)},
  title={From voxels to pixels and back: Self-supervision in natural-image reconstruction from {fMRI}},
  author={Beliy, Roman and Gaziv, Guy and Hoogi, Assaf and Strappini, Francesca and Golan, Tal and Irani, Michal},
  year={2019}
}

@article{39,
  title={Self-supervised natural image reconstruction and large-scale semantic classification from brain activity},
  author={Gaziv, Guy and Beliy, Roman and Granot, Niv and Hoogi, Assaf and Strappini, Francesca and Golan, Tal and Irani, Michal},
  journal={NeuroImage},
  volume={254},
  pages={119121},
  year={2022},
}

@inproceedings{40,
  booktitle={Advances in Neural Information Processing Systems (NeurIPS)},
  title={{Mind Reader}: Reconstructing complex images from brain activities},
  author={Lin, Sikun and Sprague, Thomas and Singh, Ambuj K},
  pages={29624--29636},
  year={2022}
}

@inproceedings{41,
  title={Reconstruction of perceived images from {fMRI} patterns and semantic brain exploration using instance-conditioned {GANs}},
  author={Ozcelik, Furkan and Choksi, Bhavin and Mozafari, Milad and Reddy, Leila and VanRullen, Rufin},
  booktitle={International Joint Conference on Neural Networks (IJCNN)},
  pages={1--8},
  year={2022},
}

@article{42,
  title={Semantics-guided hierarchical feature encoding generative adversarial network for visual image reconstruction from brain activity},
  author={Meng, Lu and Yang, Chuanhao},
  journal={IEEE Transactions on Neural Systems and Rehabilitation Engineering},
  volume={32},
  pages={1267--1283},
  year={2024},
}

@inproceedings{43,
  title={Generative adversarial networks conditioned by brain signals},
  author={Palazzo, Simone and Spampinato, Concetto and Kavasidis, Isaak and Giordano, Daniela and Shah, Mubarak},
  booktitle={Proceedings of the IEEE/CVF International Conference on Computer Vision (ICCV)},
  pages={3410--3418},
  year={2017}
}

@inproceedings{44,
  title={{Brain2Image}: Converting brain signals into images},
  author={Kavasidis, Isaak and Palazzo, Simone and Spampinato, Concetto and Giordano, Daniela and Shah, Mubarak},
  booktitle={Proceedings of the ACM International Conference on Multimedia (ACM MM)},
  pages={1809--1817},
  year={2017}
}

@inproceedings{20,
  title={High-resolution image reconstruction with latent diffusion models from human brain activity},
  author={Takagi, Yu and Nishimoto, Shinji},
  booktitle={Proceedings of the IEEE/CVF Conference on Computer Vision and Pattern Recognition (CVPR)},
  pages={14453--14463},
  year={2023}
}

@inproceedings{21,
  title={Seeing beyond the brain: Conditional diffusion model with sparse masked modeling for vision decoding},
  author={Chen, Zijiao and Qing, Jiaxin and Xiang, Tiange and Yue, Wan Lin and Zhou, Juan Helen},
  booktitle={Proceedings of the IEEE/CVF Conference on Computer Vision and Pattern Recognition (CVPR)},
  pages={22710--22720},
  year={2023}
}

@inproceedings{contrast_atten_diffuse,
  booktitle={Advances in Neural Information Processing Systems (NeurIPS)},
  title={Contrast, attend and diffuse to decode high-resolution images from brain activities},
  author={Sun, Jingyuan and Li, Mingxiao and Chen, Zijiao and Zhang, Yunhao and Wang, Shaonan and Moens, Marie-Francine},
  pages={12332--12348},
  year={2023}
}

@inproceedings{23,
  title={Controllable mind visual diffusion model},
  author={Zeng, Bohan and Li, Shanglin and Liu, Xuhui and Gao, Sicheng and Jiang, Xiaolong and Tang, Xu and Hu, Yao and Liu, Jianzhuang and Zhang, Baochang},
  booktitle={Proceedings of the AAAI Conference on Artificial Intelligence (AAAI)},
  pages={6935--6943},
  year={2024}
}

@inproceedings{24,
  booktitle={Advances in Neural Information Processing Systems (NeurIPS)},
  title={Reconstructing the mind's eye: {fMRI}-to-image with contrastive learning and diffusion priors},
  author={Scotti, Paul and Banerjee, Atmadeep and Goode, Jimmie and Shabalin, Stepan and Nguyen, Alex and Dempster, Aidan and Verlinde, Nathalie and Yundler, Elad and Weisberg, David and Norman, Kenneth and others},
  pages={24705--24728},
  year={2023}
}

@inproceedings{25,
  title={{DREAM}: Visual decoding from reversing human visual system},
  author={Xia, Weihao and de Charette, Raoul and Oztireli, Cengiz and Xue, Jing-Hao},
  booktitle={Proceedings of the IEEE/CVF Winter Conference on Applications of Computer Vision (WACV)},
  pages={8226--8235},
  year={2024}
}

@article{26,
  title={Natural scene reconstruction from {fMRI} signals using generative latent diffusion},
  author={Ozcelik, Furkan and VanRullen, Rufin},
  journal={Scientific Reports},
  volume={13},
  number={1},
  pages={15666},
  year={2023},
}

@inproceedings{27,
  title={{MindEye2}: Shared-subject models enable {fMRI}-to-image with 1 hour of data},
  author={Scotti, Paul Steven and Tripathy, Mihir and Torrico, Cesar and Kneeland, Reese and Chen, Tong and Narang, Ashutosh and Santhirasegaran, Charan and Xu, Jonathan and Naselaris, Thomas and Norman, Kenneth A. and Abraham, Tanishq Mathew},
  booktitle={International Conference on Machine Learning (ICML)},
  pages={44038--44059},
  year={2024}
}

@inproceedings{28,
  title={Brain decoding: toward real-time reconstruction of visual perception},
  author={Benchetrit, Yohann and Banville, Hubert and King, Jean-R{\'e}mi},
  booktitle={International Conference on Learning Representations (ICLR)},
  year={2024}
}

@inproceedings{45,
  title={High-resolution image synthesis with latent diffusion models},
  author={Rombach, Robin and Blattmann, Andreas and Lorenz, Dominik and Esser, Patrick and Ommer, Bj{\"o}rn},
  booktitle={Proceedings of the IEEE/CVF Conference on Computer Vision and Pattern Recognition (CVPR)},
  pages={10684--10695},
  year={2022}
}

@inproceedings{46,
  title={{DiffusionCLIP}: Text-guided diffusion models for robust image manipulation},
  author={Kim, Gwanghyun and Kwon, Taesung and Ye, Jong Chul},
  booktitle={Proceedings of the IEEE/CVF Conference on Computer Vision and Pattern Recognition (CVPR)},
  pages={2426--2435},
  year={2022}
}

@inproceedings{47,
  booktitle={Advances in Neural Information Processing Systems (NeurIPS)},
  title={Denoising diffusion probabilistic models},
  author={Ho, Jonathan and Jain, Ajay and Abbeel, Pieter},
  pages={6840--6851},
  year={2020}
}

@article{wen_neural_encoding_decoding,
  title={Neural encoding and decoding with deep learning for dynamic natural vision},
  author={Wen, Haiguang and Shi, Junxing and Zhang, Yizhen and Lu, Kun-Han and Cao, Jiayue and Liu, Zhongming},
  journal={Cerebral Cortex},
  volume={28},
  number={12},
  pages={4136--4160},
  year={2018},
}

@article{penny_for_your_thoughts,
  title={A penny for your (visual) thoughts: Self-supervised reconstruction of natural movies from brain activity},
  author={Kupershmidt, Ganit and Beliy, Roman and Gaziv, Guy and Irani, Michal},
  journal={arXiv preprint arXiv:2206.03544},
  year={2022}
}

@article{fmri_video_gan,
  title={Reconstructing rapid natural vision with {fMRI}-conditional video generative adversarial network},
  author={Wang, Chong and Yan, Hongmei and Huang, Wei and Li, Jiyi and Wang, Yuting and Fan, Yun-Shuang and Sheng, Wei and Liu, Tao and Li, Rong and Chen, Huafu},
  journal={Cerebral Cortex},
  volume={32},
  number={20},
  pages={4502--4511},
  year={2022},
}

@inproceedings{cinematic_mindscapes,
  booktitle={Advances in Neural Information Processing Systems (NeurIPS)},
  title={{Cinematic Mindscapes}: High-quality video reconstruction from brain activity},
  author={Chen, Zijiao and Qing, Jiaxin and Zhou, Juan Helen},
  pages={24841--24858},
  year={2023}
}

@article{32,
  title={Neurocine: Decoding vivid video sequences from human brain activties},
  author={Sun, Jingyuan and Li, Mingxiao and Chen, Zijiao and Moens, Marie-Francine},
  journal={arXiv preprint arXiv:2402.01590},
  year={2024}
}

@inproceedings{mishra2021eeg,
  title={{EEG} Classification for Visual Brain Decoding via Metric Learning.},
  author={Mishra, Rahul and Bhavsar, Arnav},
  booktitle={International Conference on Bioimaging (BIOIMAGING)},
  pages={160--167},
  year={2021}
}

@inproceedings{singh2023eeg2image,
  title={{EEG2Image}: Image Reconstruction from {EEG} Brain Signals},
  author={Singh, Prajwal and Pandey, Pankaj and Miyapuram, Krishna and Raman, Shanmuganathan},
  booktitle={IEEE International Conference on Acoustics, Speech and Signal Processing (ICASSP)},
  pages={1--5},
  year={2023},
}

@inproceedings{schroff2015facenet,
  title={{FaceNet}: A unified embedding for face recognition and clustering},
  author={Schroff, Florian and Kalenichenko, Dmitry and Philbin, James},
  booktitle={Proceedings of the IEEE/CVF Conference on Computer Vision and Pattern Recognition (CVPR)},
  pages={815--823},
  year={2015}
}

@inproceedings{singh2024learning,
  title={Learning robust deep visual representations from {EEG} brain recordings},
  author={Singh, Prajwal and Dalal, Dwip and Vashishtha, Gautam and Miyapuram, Krishna and Raman, Shanmuganathan},
  booktitle={Proceedings of the IEEE/CVF Winter Conference on Applications of Computer Vision (WACV)},
  pages={7553--7562},
  year={2024}
}

@inproceedings{radford2021learning,
  title={Learning transferable visual models from natural language supervision},
  author={Radford, Alec and Kim, Jong Wook and Hallacy, Chris and Ramesh, Aditya and Goh, Gabriel and Agarwal, Sandhini and Sastry, Girish and Askell, Amanda and Mishkin, Pamela and Clark, Jack and others},
  booktitle={International Conference on Machine Learning (ICML)},
  pages={8748--8763},
  year={2021},
}

@inproceedings{song2023decoding,
  title={Decoding natural images from {EEG} for object recognition},
  author={Song, Yonghao and Liu, Bingchuan and Li, Xiang and Shi, Nanlin and Wang, Yijun and Gao, Xiaorong},
  booktitle={International Conference on Learning Representations (ICLR)},
  year={2024}
}

@inproceedings{velivckovic2017graph,
  title={Graph attention networks},
  author={Veli{\v{c}}kovi{\'c}, Petar and Cucurull, Guillem and Casanova, Arantxa and Romero, Adriana and Li{\`o}, Pietro and Bengio, Yoshua},
  booktitle={International Conference on Learning Representations (ICLR)},
  year={2018}
}

@inproceedings{karras2020training,
  booktitle={Advances in Neural Information Processing Systems (NeurIPS)},
  title={Training generative adversarial networks with limited data},
  author={Karras, Tero and Aittala, Miika and Hellsten, Janne and Laine, Samuli and Lehtinen, Jaakko and Aila, Timo},
  pages={12104--12114},
  year={2020}
}

@inproceedings{liang2022mind,
  booktitle={Advances in Neural Information Processing Systems (NeurIPS)},
  title={Mind the gap: Understanding the modality gap in multi-modal contrastive representation learning},
  author={Liang, Victor Weixin and Zhang, Yuhui and Kwon, Yongchan and Yeung, Serena and Zou, James Y},
  pages={17612--17625},
  year={2022}
}

@article{zheng2015investigating,
  title={Investigating critical frequency bands and channels for {EEG}-based emotion recognition with deep neural networks},
  author={Zheng, Wei-Long and Lu, Bao-Liang},
  journal={IEEE Transactions on Autonomous Mental Development},
  volume={7},
  number={3},
  pages={162--175},
  year={2015},
}

@inproceedings{duan2013differential,
  title={Differential entropy feature for {EEG}-based emotion classification},
  author={Duan, Ruo-Nan and Zhu, Jia-Yi and Lu, Bao-Liang},
  booktitle={International IEEE/EMBS Conference on Neural Engineering (NER)},
  pages={81--84},
  year={2013},
}

@inproceedings{pumarola2021d,
  title={{D-NeRF}: Neural radiance fields for dynamic scenes},
  author={Pumarola, Albert and Corona, Enric and Pons-Moll, Gerard and Moreno-Noguer, Francesc},
  booktitle={Proceedings of the IEEE/CVF Conference on Computer Vision and Pattern Recognition (CVPR)},
  pages={10318--10327},
  year={2021}
}

@article{yao2024identifying,
  title={Identifying temporal correlations between natural single-shot videos and {EEG} signals},
  author={Yao, Yuanyuan and Stebner, Axel and Tuytelaars, Tinne and Geirnaert, Simon and Bertrand, Alexander},
  journal={Journal of Neural Engineering},
  volume={21},
  number={1},
  pages={016018},
  year={2024},
}

@article{schultz2009natural,
  title={Natural facial motion enhances cortical responses to faces},
  author={Schultz, Johannes and Pilz, Karin S},
  journal={Experimental Brain Research},
  volume={194},
  number={3},
  pages={465--475},
  year={2009},
}

@article{buccino2001action,
  title={Action observation activates premotor and parietal areas in a somatotopic manner: an {fMRI} study},
  author={Buccino, Giovanni and Binkofski, Ferdinand and Fink, Gereon R and Fadiga, Luciano and Fogassi, Leonardo and Gallese, Vittorio and Seitz, R{\"u}diger J and Zilles, Karl and Rizzolatti, Giacomo and Freund, H-J},
  journal={European Journal of Neuroscience},
  volume={13},
  number={2},
  pages={400--404},
  year={2001},
}

@inproceedings{brain_netflix,
  title={{Brain Netflix}: Scaling Data to Reconstruct Videos from Brain Signals},
  author={Fosco, Camilo and Lahner, Benjamin and Pan, Bowen and Andonian, Alex and Josephs, Emilie and Lascelles, Alex and Oliva, Aude},
  booktitle={European Conference on Computer Vision (ECCV)},
  pages={457--474},
  year={2024},
}

@article{neuralflix,
  title={{NeuroCine}: Decoding vivid video sequences from human brain activties},
  author={Sun, Jingyuan and Li, Mingxiao and Chen, Zijiao and Moens, Marie-Francine},
  journal={arXiv preprint arXiv:2402.01590},
  year={2024}
}

@inproceedings{eeg2video,
  title={{EEG2Video}: Towards decoding dynamic visual perception from {EEG} signals},
  author={Liu, Xuan-Hao and Liu, Yan-Kai and Wang, Yansen and Ren, Kan and Shi, Hanwen and Wang, Zilong and Li, Dongsheng and Lu, Bao-Liang and Zheng, Wei-Long},
  booktitle={Advances in Neural Information Processing Systems (NeurIPS)},
  pages={72245--72273},
  year={2024}
}

@inproceedings{de_cnn,
  title={Visualizing and understanding convolutional networks},
  author={Zeiler, Matthew D and Fergus, Rob},
  booktitle={European Conference on Computer Vision (ECCV)},
  pages={818--833},
  year={2014},
}

@article{buckner1998event,
  title={Event-related {fMRI} and the hemodynamic response},
  author={Buckner, Randy L},
  journal={Human Brain Mapping},
  volume={6},
  number={5-6},
  pages={373--377},
  year={1998},
}

@misc{neuroscan,
  year={2025},
  author={{Compumedics Neuroscan}},
  title={{ESI NeuroScan} System},
  howpublished={\url{https://www.compumedics.com.au/products/neuroscan/}},
  note={Accessed: 2025-03-06}
}

@misc{BioSemiActiveTwo,
  year={2025},
  author = {{BioSemi}},
  title = {{ActiveTwo {EEG} System}},
  howpublished = {\url{https://www.biosemi.com}},
  note = {Accessed: 2025-03-06}
}

@article{hartigan1979algorithm,
  title={Algorithm {AS} 136: A k-means clustering algorithm},
  author={Hartigan, John A and Wong, Manchek A},
  journal={Journal of the Royal Statistical Society. Series C (Applied Statistics)},
  volume={28},
  number={1},
  pages={100--108},
  year={1979},
}

@article{ten_twenty,
  title={The ten-twenty electrode system of the {International Federation}. {The International Federation of Clinical Neurophysiology}},
  author={Klem, George H},
  journal={Electroencephalography and Clinical Neurophysiology. Supplement},
  volume={52},
  pages={3--6},
  year={1999}
}

@article{liu2019emotion,
  title={Emotion recognition and dynamic functional connectivity analysis based on {EEG}},
  author={Liu, Xucheng and Li, Ting and Tang, Cong and Xu, Tao and Chen, Peng and Bezerianos, Anastasios and Wang, Hongtao},
  journal={IEEE Access},
  volume={7},
  pages={143293--143302},
  year={2019},
}

@article{cortes1995support,
  title={Support-vector networks},
  author={Cortes, Corinna and Vapnik, Vladimir},
  journal={Machine Learning},
  volume={20},
  pages={273--297},
  year={1995},
}

@inproceedings{kingma2014adam,
  title={Adam: A method for stochastic optimization},
  author={Kingma, Diederik P. and Ba, Jimmy},
  booktitle={International Conference on Learning Representations (ICLR)},
  year={2015}
}

@inproceedings{gong2024neuroclips,
  booktitle={Advances in Neural Information Processing Systems (NeurIPS)},
  title={{NeuroClips}: Towards high-fidelity and smooth {fMRI}-to-video reconstruction},
  author={Gong, Zixuan and Bao, Guangyin and Zhang, Qi and Wan, Zhongwei and Miao, Duoqian and Wang, Shoujin and Zhu, Lei and Wang, Changwei and Xu, Rongtao and Hu, Liang and others},
  pages={51655--51683},
  year={2024}
}

@article{dynamind2025,
  title={{DynaMind}: Reconstructing Dynamic Visual Scenes from {EEG} by Aligning Temporal Dynamics and Multimodal Semantics to Guided Diffusion},
  author={Liu, Junxiang and Lin, Junming and Li, Jiangtong and Li, Jie},
  journal={arXiv preprint arXiv:2509.01177},
  year={2025}
}

@inproceedings{jing2026evoke,
  title={{EVOKE}: Efficient and high-fidelity {EEG}-to-video reconstruction via decoupling implicit neural representation},
  author={Jing, Haodong and Yang, Panqi and Jiang, Dongyao and Liu, Zhipeng and Zheng, Nanning and Ma, Yongqiang},
  booktitle={Proceedings of the AAAI Conference on Artificial Intelligence (AAAI)},
  pages={5539--5547},
  year={2026}
}

@article{mindcine2026,
  title={{MindCine}: Multimodal {EEG}-to-Video Reconstruction with Large-Scale Pretrained Models},
  author={Zhou, Tian-Yi and Liu, Xuan-Hao and Lu, Bao-Liang and Zheng, Wei-Long},
  journal={arXiv preprint arXiv:2601.18192},
  year={2026}
}

@inproceedings{han2026cross,
  title={Cross-Subject {EEG}-to-Video Reconstruction and Beyond},
  author={Han, Runduo and Tan, Hongchen},
  booktitle={Proceedings of the IEEE/CVF Conference on Computer Vision and Pattern Recognition (CVPR)},
  pages={23294--23303},
  year={2026}
}

@inproceedings{eegmirror2025,
  title={{EEGMirror}: Leveraging {EEG} Data in the Wild via Montage-Agnostic Self-Supervision for {EEG} to Video Decoding},
  author={Liu, Xuan-Hao and Lu, Bao-Liang and Zheng, Wei-Long},
  booktitle={Proceedings of the IEEE/CVF International Conference on Computer Vision (ICCV)},
  pages={18273--18283},
  year={2025}
}

@inproceedings{need2025,
  title={{NEED}: Cross-Subject and Cross-Task Generalization for Video and Image Reconstruction from {EEG} Signals},
  author={Huang, Shuai and Luo, Huan and Jing, Haodong and Zhang, Qixian and Chang, Litao and Feng, Yating and Lin, Xiao and Qin, Chendong and Chen, Han and Jia, Shuwen and Sun, Siyi and Wang, Yongxiong},
  booktitle={Advances in Neural Information Processing Systems (NeurIPS)},
  year={2025}
}

@inproceedings{wang2019multi,
  title={Multi-Similarity Loss With General Pair Weighting for Deep Metric Learning},
  author={Wang, Xun and Han, Xintong and Huang, Weilin and Dong, Dengke and Scott, Matthew R.},
  booktitle={Proceedings of the IEEE/CVF Conference on Computer Vision and Pattern Recognition (CVPR)},
  pages={5022--5030},
  year={2019}
}
\bibliographystyle{tmlr}

\clearpage
\appendix
\section{Datasets}
\label{sec:appendix_datasets}

We use three EEG-video datasets. The first is SEED (SJTU Emotion EEG Dataset) \citep{duan2013differential, zheng2015investigating}, an emotion-based dataset. It consists of EEG recordings from $15$ participants ($7$ males and $8$ females), collected over three sessions spaced approximately one week apart. In each session, participants watched $15$ video clips of about four minutes each, selected to induce three emotional states: positive, neutral, and negative. EEG was recorded with a $62$-channel ESI NeuroScan System \citep{neuroscan}.

The second is the EEG-Video Action dataset \citep{yao2024identifying}, designed to study the neural representation of natural single-shot videos. It comprises EEG from $19$ participants with normal or corrected-to-normal vision while they watched $13$ muted YouTube videos. EEG was recorded with a $64$-channel BioSemi ActiveTwo \citep{BioSemiActiveTwo} system at $2048$\,Hz.

The third is SEED-DV \citep{eeg2video}. It contains EEG from $20$ subjects recorded while viewing $1400$ video clips across $40$ concepts.

\section{Analysis of SEED-DV}
\label{sec:appendix_seedandseeddv}

We follow the procedure of \citet{eeg2video} for SEED-DV. A triplet loss conditioned on the video classes is used to learn EEG features. We adopt the $7$-fold protocol of \citet{eeg2video}, using $5$ blocks for training, $1$ for validation, and $1$ for testing. Linear probing with an SVM~\citep{cortes1995support} is used for a fair comparison with the EEG2Video feature extractor.

Using the GLMNet pipeline~\citep{eeg2video}, raw EEG achieves an average $40$-way classification accuracy of $6.20\%$ across $20$ subjects in our reproduction. Our EEG-only model achieves $2.9\%$ with linear probing, averaged over subjects (chance: $2.5\%$). Its $k$-means accuracy is $25.6\%$, rising to $26.3\%$ when text features from a pre-trained CLIP~\citep{radford2021learning} model are added with the EEG network unchanged. The two metrics differ in protocol. Linear probing trains a classifier on the training blocks and tests on a held-out block. $K$-means accuracy clusters the embeddings without labels and matches clusters to classes, so the two numbers are not directly comparable. Our linear-probe accuracy is below that of GLMNet, so we use SEED-DV as an analysis benchmark rather than as a reconstruction benchmark.

\begin{figure*}[tp]
    \centering
    \includegraphics[width=1.0\linewidth]{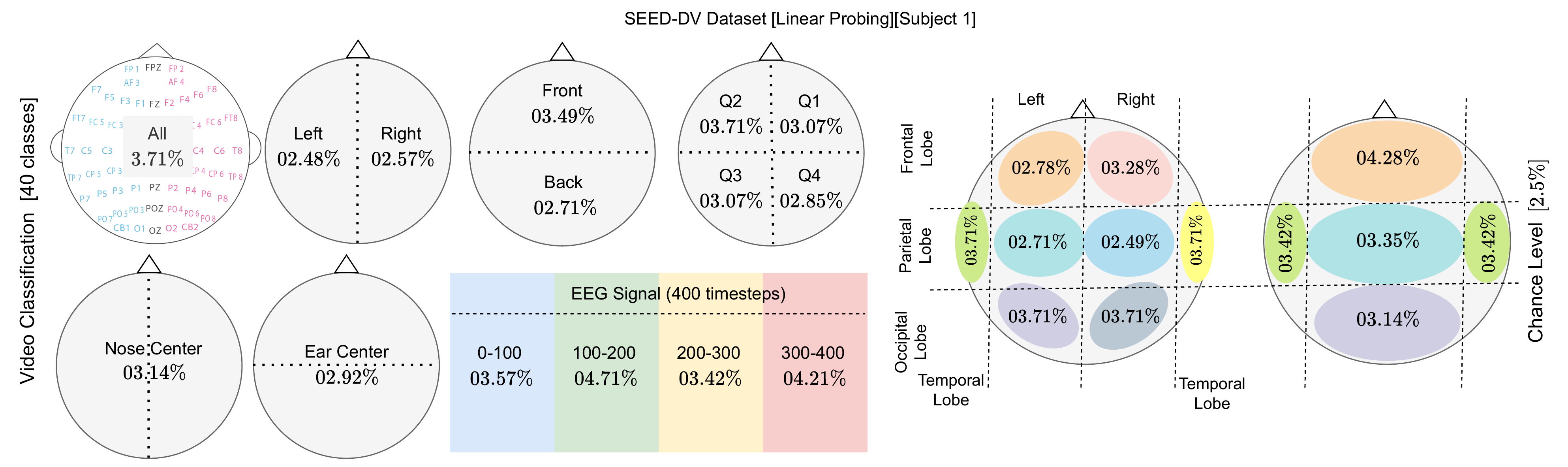}
    \caption{\textbf{SEED-DV Brain Region.} Region-wise linear-probing accuracy for subject $1$ of the SEED-DV dataset \citep{eeg2video}. The EEG channels are sampled from each specific brain region, and the representation is learned using a triplet-loss-based method. Linear probing is used for consistency with EEG2Video \citep{eeg2video}.}
    \label{fig:seed_dv_topomap}\vspace{-0.5em}
\end{figure*}

SEED-DV is challenging because of its many fine-grained classes and few repetitions per class. The region-wise and temporal analyses in Figure~\ref{fig:seed_dv_topomap} stay close to chance, so they show no reliable regional or temporal differences.

\section{Emotion Classification on SEED}
\label{sec:appendix_seedemotion}

\begin{figure*}[!t]
    \centering
    \includegraphics[width=1.0\linewidth]{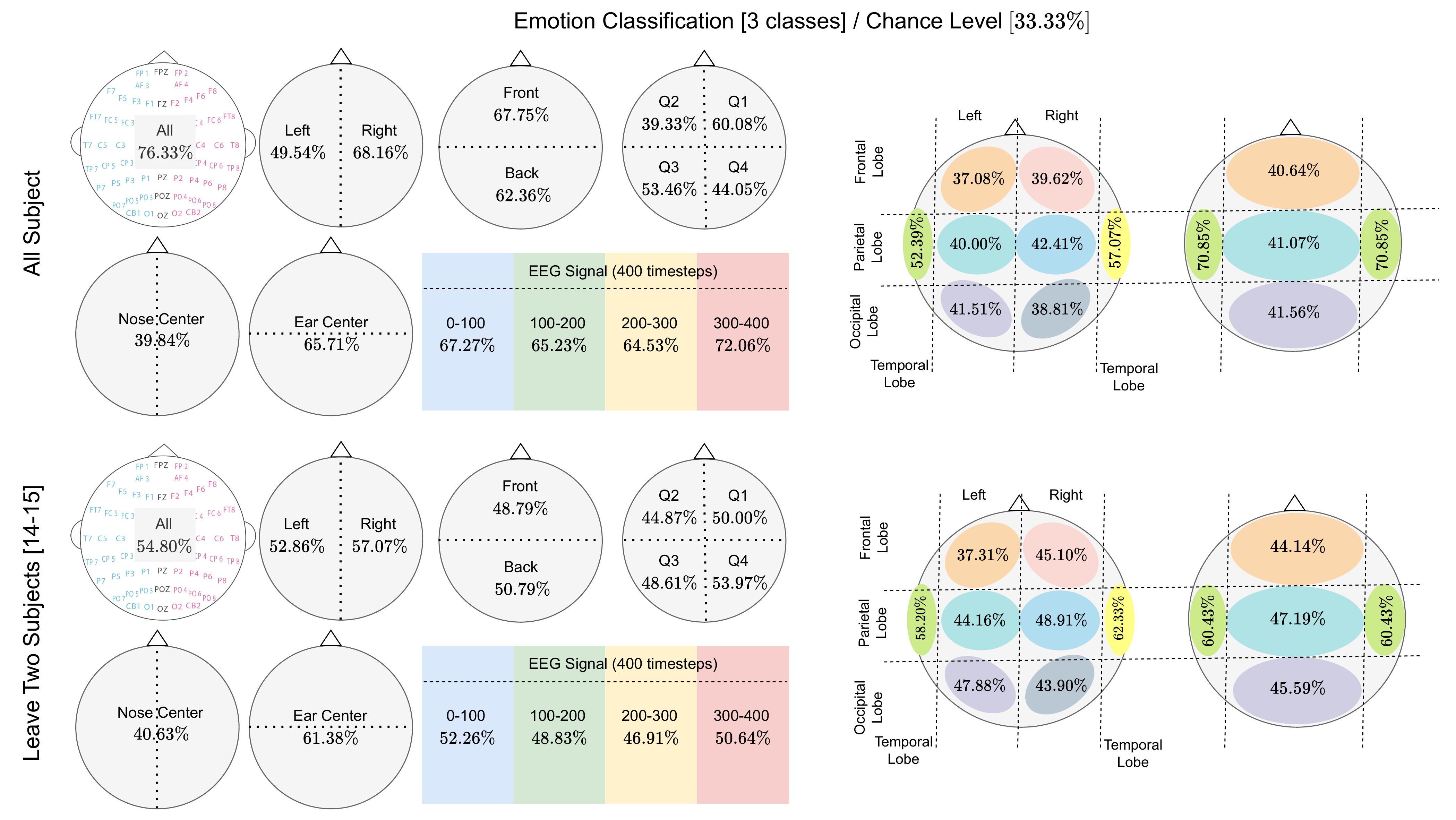}
    \caption{\textbf{Emotion Decoding on SEED}. The figure shows the test classification accuracy of EEG signals when trained on different brain region channels for emotion classification on the SEED dataset \citep{duan2013differential, zheng2015investigating}. The two rows show the two evaluation regimes. Under All Subjects, training, validation and test segments come from the same recordings but do not overlap. Under Leave Two Subjects, the network is trained on the data of subjects $1$ to $13$ and tested on subjects $14$ and $15$ from the SEED dataset. To show the effect of different non-overlapping time segments in the case of a pre-trained network, we mask one $100$-timestep window and compute the accuracy with all EEG channels.}
    \label{fig:seed_emotion}
    \vspace{-1.0em}
\end{figure*}

We classify emotion into three categories (positive, neutral, and negative). As shown in Figure~\ref{fig:seed_emotion}, the model reaches $76.33\%$ under the All Subjects regime (chance $33.33\%$). Unlike video classification, the Right Hemisphere outperforms the Left ($68.16\%$ vs.\ $49.54\%$). The Front Region outperforms the Back Region ($67.75\%$ vs.\ $62.36\%$). For reference, \citet{liu2019emotion} report $87.03\%$ on SEED. We do not optimize for emotion decoding. Our goal is to compare information content across regions and time, not to reach state-of-the-art emotion recognition.

Under the Leave Two Subjects regime, accuracy drops to $54.80\%$. The relative drop is smaller than for video classification. The Right Hemisphere still outperforms the Left ($57.07\%$ vs.\ $52.86\%$).

\section{Implementation Details}
\label{sec:appendix_implement}

\textbf{Pre-processing.} For SEED, we use the pre-processed EEG released by the authors, downsampled to $200$\,Hz and band-pass filtered at $0$--$75$\,Hz. For EEG-Video Action, we follow the pre-processing of \citet{yao2024identifying}: a) interpolate bad channels, b) re-reference to the average, c) apply a notch filter to remove line noise, d) apply a high-pass filter at $0.5$\,Hz, e) downsample to $200$\,Hz, and f) regress out the electrooculography (EOG) channels to remove eye-movement artifacts.

For SEED and EEG-Video Action, we cut the EEG and the corresponding video into non-overlapping $2$-second segments of $400$ time points. For each video, $80\%$ of its segments are randomly assigned to training, $10\%$ to validation, and $10\%$ to testing. In the Leave Two Subjects regime, all segments of the last two subjects form the test split (subjects $14$ and $15$ for SEED, $18$ and $19$ for EEG-Video Action), and the remaining segments form the training split.

\vspace{1mm} \noindent
\textbf{EEG Feature Extraction.} We use the NICE \citep{song2023decoding} architecture as the backbone, producing a unit-normalized feature $\psi \in \mathbb{R}^{1024}$. We train it with the triplet loss \citep{schroff2015facenet} and a multi-similarity miner~\citep{wang2019multi}, which selects hard and semi-hard triplets. We use the Adam optimizer \citep{kingma2014adam} with learning rate $3\times10^{-4}$ and weight decay $10^{-4}$, for $300$ epochs.

\begin{lstlisting}[language=Python, caption=Sinusoidal positional encoding and conditioning vector, label=code:posenc]
import numpy as np
import torch

def encode_position_vectorized(x, enc_dim):
    # Frequencies [2^0, 2^1, ..., 2^(enc_dim-1)]
    powers = torch.arange(enc_dim, device=x.device)
    frequencies = torch.pow(2.0, powers)
    xf = x.unsqueeze(-1) * frequencies
    # Interleave sin and cos values
    encodings = torch.stack([torch.sin(xf), torch.cos(xf)], dim=-1).flatten(-2)
    # Prepend the raw index
    return torch.cat([x.unsqueeze(-1), encodings], dim=-1)

images_path, images_idx, labels, cond_feats = [], [], [], []
for eeg_feat, class_id, video_frames_path in x_dataset:
    eeg_feat = eeg_feat.detach().cpu()
    for per_frame_idx in range(total_frames):
        images_path.append(video_frames_path)
        images_idx.append(per_frame_idx)
        labels.append(class_id * total_frames + per_frame_idx)
        pe = encode_position_vectorized(torch.tensor(per_frame_idx).float(), 10)
        cond_feats.append(torch.cat([eeg_feat, pe], dim=-1))

cond_feats = torch.stack(cond_feats, dim=0).to(torch.float32)
labels     = torch.from_numpy(np.array(labels)).to(torch.int32)
images_idx = torch.from_numpy(np.array(images_idx)).to(torch.int32)
\end{lstlisting}

We train the GAN for $5500$ kimg (thousands of real images shown to the discriminator) on SEED \citep{duan2013differential,zheng2015investigating} and EEG-Video Action \citep{yao2024identifying}, and for $2500$ kimg on SEED-DV \citep{eeg2video}, with the following hyperparameters:

\begin{lstlisting}[language=Python, caption=Hyperparameters for EEGStyleGAN-ADA, label=code:videogan]
# SEED and EEG-Video Action
'cifar': dict(ref_gpus=1, kimg=5500, mb=32, mbstd=8, fmaps=0.75, lrate=0.002, gamma=5, ema=20, ramp=0.05, map=8)

# SEED-DV
'cifar': dict(ref_gpus=1, kimg=2500, mb=16, mbstd=4, fmaps=1.0, lrate=0.0015, gamma=2, ema=10, ramp=0.05, map=4)
\end{lstlisting}

\section{Generation Details and Temporal Ablation}
\label{sec:appendix_generation}

We synthesize clips of $8$ frames per $2$-second segment. Figure~\ref{fig:teaser}(a) shows qualitative examples. Generated frames capture coarse semantic structure and motion patterns of the viewed videos. Fine spatial detail comes from the dataset-specific generative prior, while the EEG provides semantic guidance.

We evaluate reconstruction quality with PSNR, SSIM, and LPIPS, and semantic alignment with CLIP similarity between generated and ground-truth frames. Standard video metrics such as FVD are unsuitable here because the clips are short and highly redundant.

\vspace{1mm} \noindent
\textbf{EEG Feature to Video.} We use EEGStyleGAN-ADA \citep{singh2024learning}, which modifies the conditioning network of StyleGAN-ADA \citep{karras2020training} to synthesize images from EEG. We additionally condition it on a sinusoidal positional encoding of the frame index (Listing~\ref{code:posenc}), which lets an image generator produce the frames of a video. We use $10$ frequencies ($21$-d encoding) for SEED and SEED-DV because this gave better perceptual quality than lower values. The EEG-Video Action model, used for all EEG-Video Action results and the controlled diagnostics, uses $2$ frequencies ($5$-d). Listing~\ref{code:posenc} shows the $10$-frequency setting.

\begin{figure*}[!tp]
    \centering
    \includegraphics[width=0.85\linewidth]{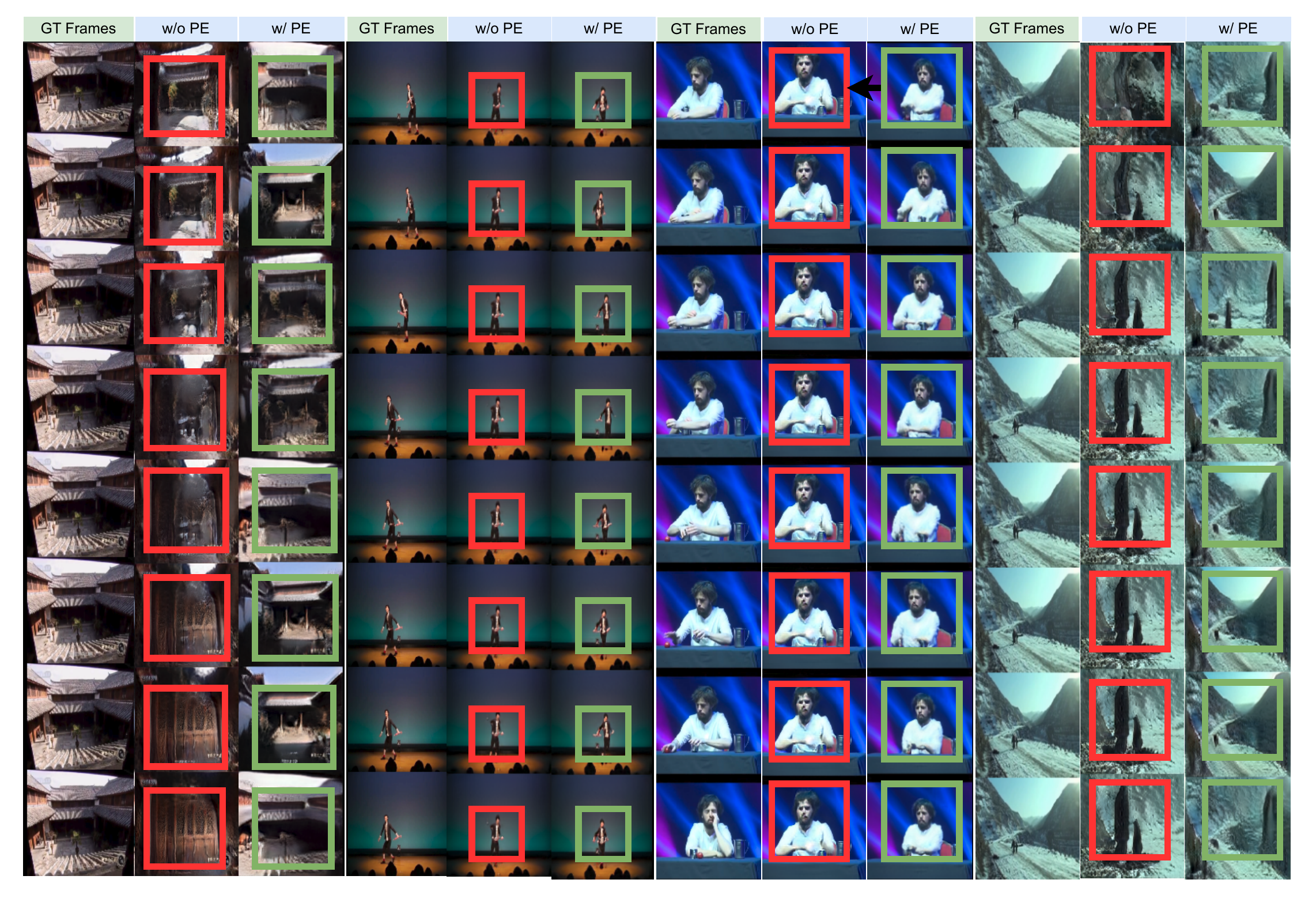}\vspace{-0.5em}
    \caption{\textbf{Temporal Ablation.} Qualitative comparison of networks trained without and with positional encoding (PE) of the frame index. Without PE, the generator produces static or interpolated frames (red boxes), while with PE, it produces structured temporal dynamics such as camera or object motion (green boxes).  This shows the role of temporal conditioning in enabling dynamic video generation from EEG.}
    \label{fig:temporalabl}
\end{figure*}

\vspace{1mm} \noindent
\textbf{Effect of Temporal Conditioning.} Without positional encoding (PE), the generator has no explicit frame index. It produces frames with much less motion or degenerate fading over time. With PE, it produces temporally varying frames (Figure~\ref{fig:temporalabl}). Table~\ref{table:pe_ablation} quantifies this with an optical-flow score (OFS). Without PE, OFS stays well below that of ground-truth video, and it increases substantially with PE. With PE, OFS exceeds that of ground-truth video, by about $1.9\times$ on EEG-Video Action and $3\times$ on SEED, so part of the generated motion is stronger than real motion. Qualitative inspection shows that this motion is structured rather than noise. The frame index thus controls temporal variation, while the EEG controls content.

\begin{table}[!t]
\centering
\caption{\textbf{Positional Encoding Ablation.} Quantitative comparison of Optical Flow Score (OFS) across the video frames when generated with and without positional encoding over the frame index.}\vspace{-0.5em}
\resizebox{0.8\linewidth}{!}{
\def\arraystretch{1.5}
\begin{tabular}{c|c|c|c}
\hline
Dataset & OFS w/o PE & OFS w/ PE & OFS GT Video \\ \hline
SEED \citep{duan2013differential, zheng2015investigating} & 0.561 & 4.181 & 1.371\\
EEG-Video Action \citep{yao2024identifying} & 0.276 & 0.896 & 0.479 \\
SEED-DV \citep{eeg2video} & 0.260 & 2.246 & 1.443 \\ \hline
\end{tabular}}
\label{table:pe_ablation}
\end{table}

\section{Class-Conditional Baseline: Follow-up Checks}
\label{sec:appendix_classcond}

\textbf{Diversity check.} For twenty randomly sampled class-and-position cells, we generate eight frames from eight independent noise draws with the conditioning fixed. We compute the mean pairwise LPIPS (AlexNet) among these generated frames and among the real segments that share the same cell. Generated diversity averages $0.258$ against a true diversity of $0.265$ (ratio $0.97$). The class-conditional model therefore does not collapse to a single output per cell.

\begin{figure}[t]
\centering
\includegraphics[width=0.99\linewidth]{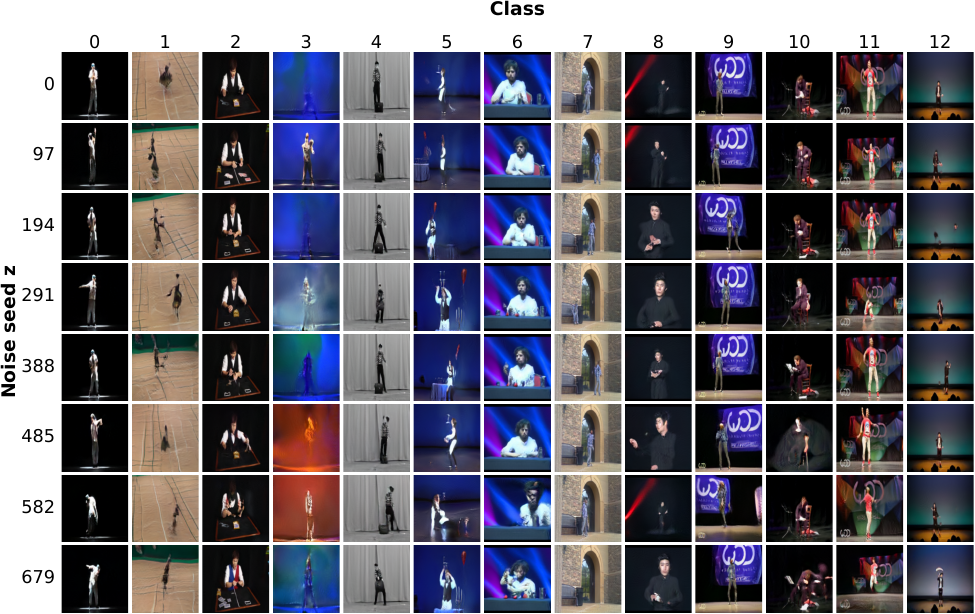}
\caption{\textbf{Noise-consistency grid} (class-conditional model, frame position 4). Rows share a noise draw; columns share a class. Content changes across columns and is stable within a column. No shared trait is visible along a row.}
\label{fig:zgrid}
\end{figure}

\vspace{1mm} \noindent
\textbf{Noise-consistency check.} We also test whether a fixed noise draw carries a visual trait that persists across classes. Each generated frame is reduced to a coarse layout descriptor: the channel-mean grayscale image, area-downsampled to $16\times16$ and normalized to zero mean and unit variance. We compare $\ell_2$ descriptor distances for three kinds of pairs. Same-noise, different-class pairs average $21.18$, no closer than a shuffled baseline of random pairs ($20.58$). Different-noise, same-class pairs average $13.24$. Class conditioning thus determines scene layout, while a fixed noise draw carries no detectable shared trait across classes (Figure~\ref{fig:zgrid}).

\section{Generalization Across Time}
\label{sec:appendix_time}

\begin{figure*}[tp]
    \centering
    \includegraphics[width=\linewidth]{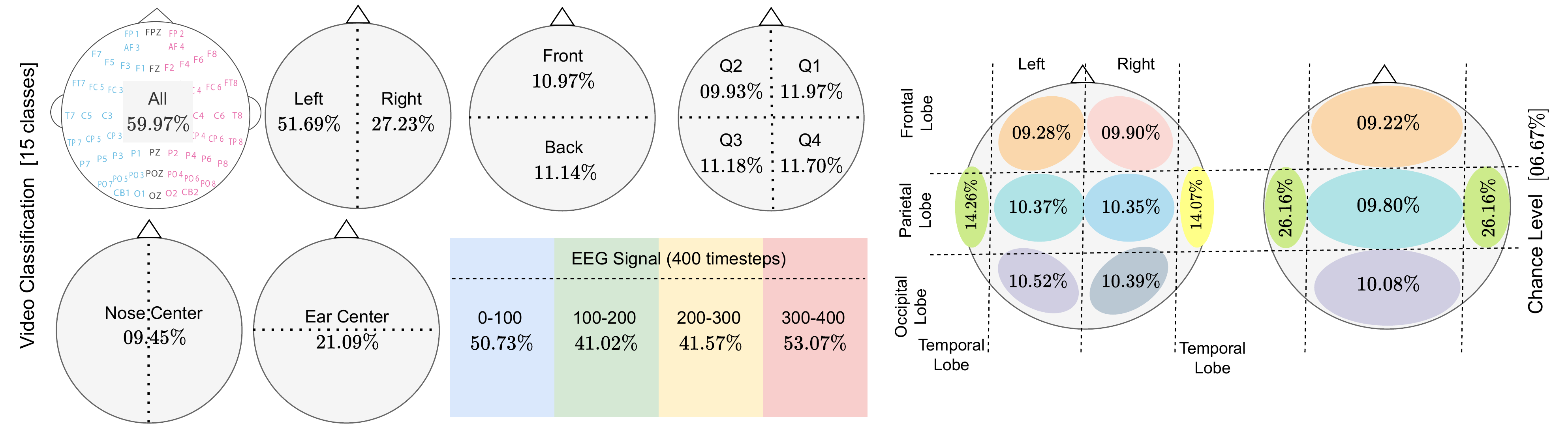}
    \caption{\textbf{Generalization across time} (SEED \citep{duan2013differential, zheng2015investigating}). The first $78\%$ of the timesteps are used for training and the last two $10\%$ blocks for validation and testing, with no overlapping segments. Region-wise and time-masked video classification accuracy on the held-out later timesteps.}
    \label{fig:seed_leavetime}
\end{figure*}

We split the SEED \citep{duan2013differential, zheng2015investigating} EEG by time to test whether the representation generalizes to later, unseen portions of the recording. The first $78\%$ of the timesteps are used for training and the last two $10\%$ blocks for validation and testing. The $2\%$ between them is discarded so that no training segment overlaps validation or test. As shown in Figure~\ref{fig:seed_leavetime}, video classification reaches $59.97\%$ on the held-out later timesteps (chance $6.67\%$). The left hemisphere again outperforms the right ($51.69\%$ vs.\ $27.23\%$).

\section{EEG Cap Montage}
\label{sec:appendix_section_eeg_montage_cap}

\begin{figure*}[!t]
\centering
\begin{minipage}[b]{0.45\linewidth}
  \centering
  \centerline{\includegraphics[width=\linewidth]{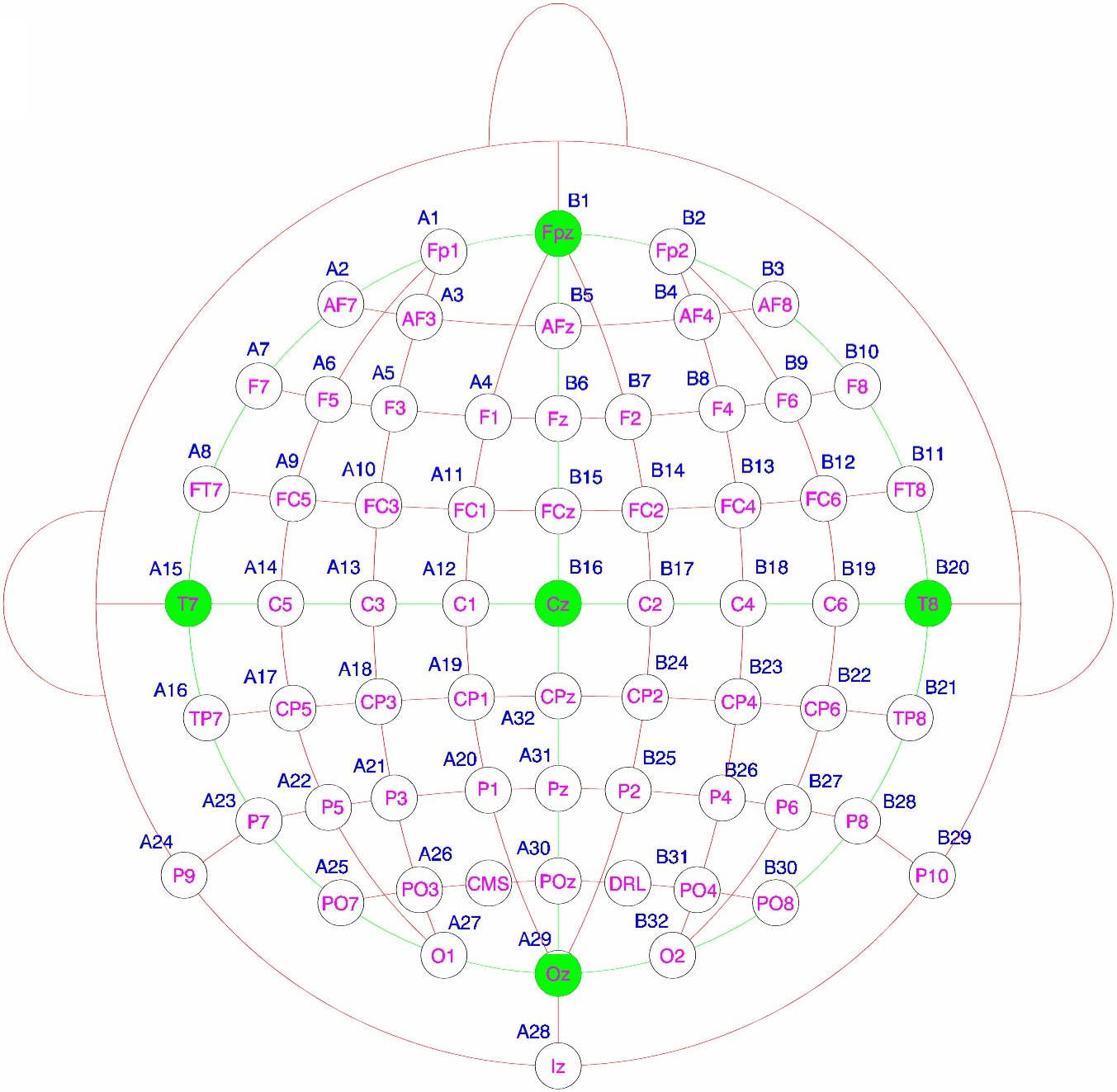}}
  \centerline{\small (a)}\medskip
\end{minipage}
\begin{minipage}[b]{.45\linewidth}
  \centering
  \centerline{\includegraphics[width=\linewidth]{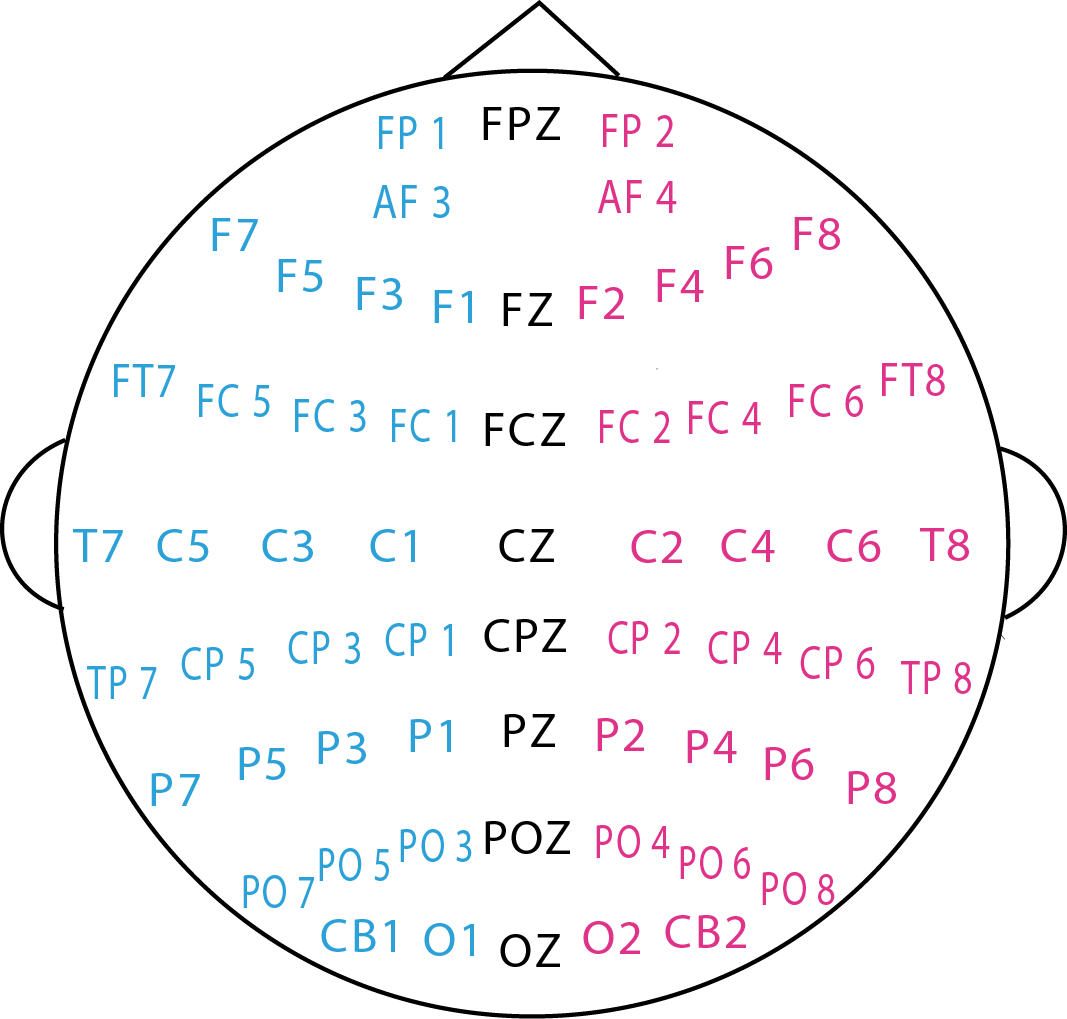}}
  \centerline{\small (b)}\medskip
\end{minipage}
\caption{\textbf{EEG montages.} Electrode layouts used to select the channels of each brain region for (a) EEG-Video Action \citep{yao2024identifying} and (b) SEED \citep{duan2013differential, zheng2015investigating}.}
\label{fig:eeg_cap_montage}
\end{figure*}

Figure~\ref{fig:eeg_cap_montage} shows the electrode layouts for the EEG-Video Action and SEED datasets. Both follow the international $10$--$20$ system~\citep{ten_twenty}. The listings below give the channel indices used to select each brain region on each cap.

\begin{lstlisting}[language=Python, caption=SEED dataset EEG cap locations, label=code:seedcap]
EEG_CAP_SEED_V1 = {
'all': np.array([1, 2, 3, 4, 5, 6, 7, 8, 9, 10, 11, 12, 13, 14, 15, 16, 17, 18, 19, 20, 21, 22, 23, 24, 25, 26, 27, 28, 29, 30, 31, 32, 33, 34, 35, 36, 37, 38, 39, 40, 41, 42, 43, 44, 45, 46, 47, 48, 49, 50, 51, 52, 53, 54, 55, 56, 57, 58, 59, 60, 61, 62], dtype=np.int32),
'noseback_left': np.array([1, 4, 6, 7, 8, 9, 15, 16, 17, 18, 24, 25, 26, 27, 33, 34, 35, 36, 42, 43, 44, 45, 51, 52, 53, 58, 59], dtype=np.int32),
'noseback_center': np.array([2, 10, 19, 28, 37, 46, 54, 60], dtype=np.int32),
'noseback_right': np.array([3, 5, 11, 12, 13, 14, 20, 21, 22, 23, 29, 30, 31, 32, 38, 39, 40, 41, 47, 48, 49, 50, 55, 56, 57, 61, 62], dtype=np.int32),
'noseback_Q1_right': np.array([3, 5, 11, 12, 13, 14, 20, 21, 22, 23], dtype=np.int32),
'noseback_Q2_left': np.array([1, 4, 6, 7, 8, 9, 15, 16, 17, 18], dtype=np.int32),
'noseback_Q3_left': np.array([33, 34, 35, 36, 42, 43, 44, 45, 51, 52, 53, 58, 59], dtype=np.int32),
'noseback_Q4_right': np.array([38, 39, 40, 41, 47, 48, 49, 50, 55, 56, 57, 61, 62], dtype=np.int32),
'leftrightear_center': np.array([24, 25, 26, 27, 28, 29, 30, 31, 32], dtype=np.int32),
'noseback_front': np.array([1, 2, 3, 4, 5, 6, 7, 8, 9, 10, 11, 12, 13, 14, 15, 16, 17, 18, 19, 20, 21, 22, 23], dtype=np.int32),
'noseback_back': np.array([33, 34, 35, 36, 37, 38, 39, 40, 41, 42, 43, 44, 45, 46, 47, 48, 49, 50, 51, 52, 53, 54, 55, 56, 57, 58, 59, 60, 61, 62], dtype=np.int32),
'lobes_frontal_left': np.array([1, 4, 6, 7, 8, 9, 16, 17, 18], dtype=np.int32),
'lobes_frontal_right': np.array([3, 5, 11, 12, 13, 14, 20, 21, 22], dtype=np.int32),
'lobes_parietal_left': np.array([34, 35, 36, 42, 43, 44, 45], dtype=np.int32),
'lobes_parietal_right': np.array([36, 39, 40, 47, 48, 49, 50], dtype=np.int32),
'lobes_occipital_left': np.array([51, 52, 53, 59], dtype=np.int32),
'lobes_occipital_right': np.array([55, 56, 57, 61], dtype=np.int32),
'lobes_temporal_left': np.array([15, 24, 33], dtype=np.int32),
'lobes_temporal_right': np.array([23, 32, 41], dtype=np.int32),
'lobes_frontal': np.array([1, 4, 6, 7, 8, 9, 16, 17, 18, 3, 5, 11, 12, 13, 14, 20, 21, 22, 2, 10, 19], dtype=np.int32),
'lobes_parietal': np.array([34, 35, 36, 42, 43, 44, 45, 36, 39, 40, 47, 48, 49, 50, 37, 46], dtype=np.int32),
'lobes_occipital': np.array([51, 52, 53, 59, 55, 56, 57, 61, 54, 60], dtype=np.int32),
'lobes_temporal': np.array([15, 23, 24, 32, 33, 41], dtype=np.int32),
}
\end{lstlisting}

\begin{lstlisting}[language=Python, caption=EEG-Video Action dataset EEG cap locations, label=code:actioncap]
EEG_CAP_SHOT_V1 = {
'all': np.array([1, 2, 3, 4, 5, 6, 7, 8, 9, 10, 11, 12, 13, 14, 15, 16, 17, 18, 19, 20, 21, 22, 23, 24, 25, 26, 27, 28, 29, 30, 31, 32, 33, 34, 35, 36, 37, 38, 39, 40, 41, 42, 43, 44, 45, 46, 47, 48, 49, 50, 51, 52, 53, 54, 55, 56, 57, 58, 59, 60, 61, 62, 63, 64], dtype=np.int32),
'noseback_left': np.array([1, 2, 3, 7, 6, 5, 4, 8, 9, 10, 11, 15, 14, 13, 12, 16, 17, 18, 19, 24, 23, 22, 21, 20, 25, 26, 27], dtype=np.int32),
'noseback_center': np.array([33, 37, 38, 47, 48, 32, 31, 30, 29, 28], dtype=np.int32),
'noseback_right': np.array([34, 36, 35, 39, 40, 41, 42, 46, 45, 44, 43, 49, 50, 51, 52, 56, 55, 54, 53, 57, 58, 59, 60, 61, 63, 62, 64], dtype=np.int32),
'noseback_Q1_right': np.array([34, 36, 35, 39, 40, 41, 42, 46, 45, 44, 43], dtype=np.int32),
'noseback_Q2_left': np.array([1, 2, 3, 7, 6, 5, 4, 8, 9, 10, 11], dtype=np.int32),
'noseback_Q3_left': np.array([16, 17, 18, 19, 24, 23, 22, 21, 20, 25, 26, 27], dtype=np.int32),
'noseback_Q4_right': np.array([56, 55, 54, 53, 57, 58, 59, 60, 61, 63, 62, 64], dtype=np.int32),
'leftrightear_center': np.array([15, 14, 13, 12, 48, 49, 50, 51, 52], dtype=np.int32),
'noseback_front': np.array([1, 33, 34, 2, 3, 37, 36, 35, 7, 6, 5, 4, 38, 39, 40, 41, 42, 8, 9, 10, 11, 47, 46, 45, 44, 43], dtype=np.int32),
'noseback_back': np.array([16, 17, 18, 19, 32, 56, 55, 54, 53, 24, 23, 22, 21, 20, 31, 57, 58, 59, 60, 61, 25, 26, 30, 63, 62, 27, 29, 64, 28], dtype=np.int32),
'lobes_frontal_left': np.array([1, 2, 3, 4, 5, 6, 7, 9, 10, 11], dtype=np.int32),
'lobes_frontal_right': np.array([34, 35, 36, 39, 40, 41, 42, 46, 45, 44], dtype=np.int32),
'lobes_parietal_left': np.array([17, 18, 19, 24, 23, 22, 21, 20], dtype=np.int32),
'lobes_parietal_right': np.array([56, 55, 54, 57, 58, 59, 60, 61], dtype=np.int32),
'lobes_occipital_left': np.array([25, 26, 27], dtype=np.int32),
'lobes_occipital_right': np.array([63, 62, 64], dtype=np.int32),
'lobes_temporal_left': np.array([8, 15, 16], dtype=np.int32),
'lobes_temporal_right': np.array([43, 52, 53], dtype=np.int32),
'lobes_frontal': np.array([1, 2, 3, 4, 5, 6, 7, 9, 10, 11, 34, 35, 36, 39, 40, 41, 42, 46, 45, 44, 33, 37, 38, 47], dtype=np.int32),
'lobes_parietal': np.array([17, 18, 19, 24, 23, 22, 21, 20, 56, 55, 54, 57, 58, 59, 60, 61, 32, 31], dtype=np.int32),
'lobes_occipital': np.array([25, 26, 27, 63, 62, 64, 30, 29], dtype=np.int32),
'lobes_temporal': np.array([8, 15, 16, 43, 52, 53], dtype=np.int32),
}
\end{lstlisting}

\begin{lstlisting}[language=Python, caption=SEED-DV dataset EEG cap locations, label=code:seeddvncap]
SEEDV_CAP = {
'all': np.array([1, 2, 3, 4, 5, 6, 7, 8, 9, 10, 11, 12, 13, 14, 15, 16, 17, 18, 19, 20, 21, 22, 23, 24, 25, 26, 27, 28, 29, 30, 31, 32, 33, 34, 35, 36, 37, 38, 39, 40, 41, 42, 43, 44, 45, 46, 47, 48, 49, 50, 51, 52, 53, 54, 55, 56, 57, 58, 59, 60, 61], dtype=np.int32),
'noseback_left': np.array([1, 4, 6, 7, 8, 9, 15, 16, 17, 18, 24, 25, 26, 27, 33, 34, 35, 36, 42, 43, 44, 45, 51, 52, 53, 58, 59], dtype=np.int32),
'noseback_center': np.array([2, 10, 19, 28, 37, 46, 54, 60], dtype=np.int32),
'noseback_right': np.array([3, 5, 11, 12, 13, 14, 20, 21, 22, 23, 29, 30, 31, 32, 38, 39, 40, 41, 47, 48, 49, 50, 55, 56, 57, 61, 62], dtype=np.int32),
'noseback_Q1_right': np.array([3, 5, 11, 12, 13, 14, 20, 21, 22, 23], dtype=np.int32),
'noseback_Q2_left': np.array([1, 3, 6, 7, 8, 9, 15, 16, 17, 18], dtype=np.int32),
'noseback_Q3_left': np.array([33, 34, 35, 36, 42, 43, 44, 45, 51, 52, 53, 56, 59], dtype=np.int32),
'noseback_Q4_right': np.array([38, 39, 40, 41, 47, 48, 49, 50, 55, 56, 57, 61, 62], dtype=np.int32),
'leftrightear_center': np.array([24, 25, 26, 27, 28, 29, 30, 31, 32], dtype=np.int32),
'noseback_front': np.array([1, 2, 3, 4, 5, 6, 7, 8, 9, 10, 11, 12, 13, 14, 15, 16, 17, 18, 19, 20, 21, 22, 23], dtype=np.int32),
'noseback_back': np.array([33, 34, 35, 36, 37, 38, 39, 40, 41, 42, 43, 44, 45, 46, 47, 48, 49, 50, 51, 52, 53, 54, 55, 56, 57, 58, 59, 60, 61, 62], dtype=np.int32),
'lobes_frontal_left': np.array([1, 4, 6, 7, 8, 9, 16, 17, 18], dtype=np.int32),
'lobes_frontal_right': np.array([3, 5, 11, 12, 13, 14, 20, 21, 22], dtype=np.int32),
'lobes_parietal_left': np.array([34, 35, 36, 42, 43, 44, 45], dtype=np.int32),
'lobes_parietal_right': np.array([38, 39, 40, 47, 48, 49, 50], dtype=np.int32),
'lobes_occipital_left': np.array([51, 52, 53, 54], dtype=np.int32),
'lobes_occipital_right': np.array([55, 56, 57, 61], dtype=np.int32),
'lobes_temporal_left': np.array([15, 29, 33], dtype=np.int32),
'lobes_temporal_right': np.array([23, 32, 41], dtype=np.int32),
'lobes_frontal': np.array([1, 2, 3, 4, 5, 6, 7, 8, 9, 10, 11, 12, 13, 14, 16, 17, 18, 19, 20, 21, 22], dtype=np.int32),
'lobes_parietal': np.array([34, 35, 36, 37, 38, 39, 40, 42, 43, 44, 45, 46, 47, 48, 49, 50], dtype=np.int32),
'lobes_occipital': np.array([51, 52, 53, 54, 55, 56, 57, 59, 60, 61], dtype=np.int32),
'lobes_temporal': np.array([15, 29, 33, 23, 32, 41], dtype=np.int32),
}
\end{lstlisting}

\section{Additional Qualitative Results}
\label{sec:appendix_qual}

\begin{figure*}[tp]
    \centering
    \includegraphics[width=\linewidth]{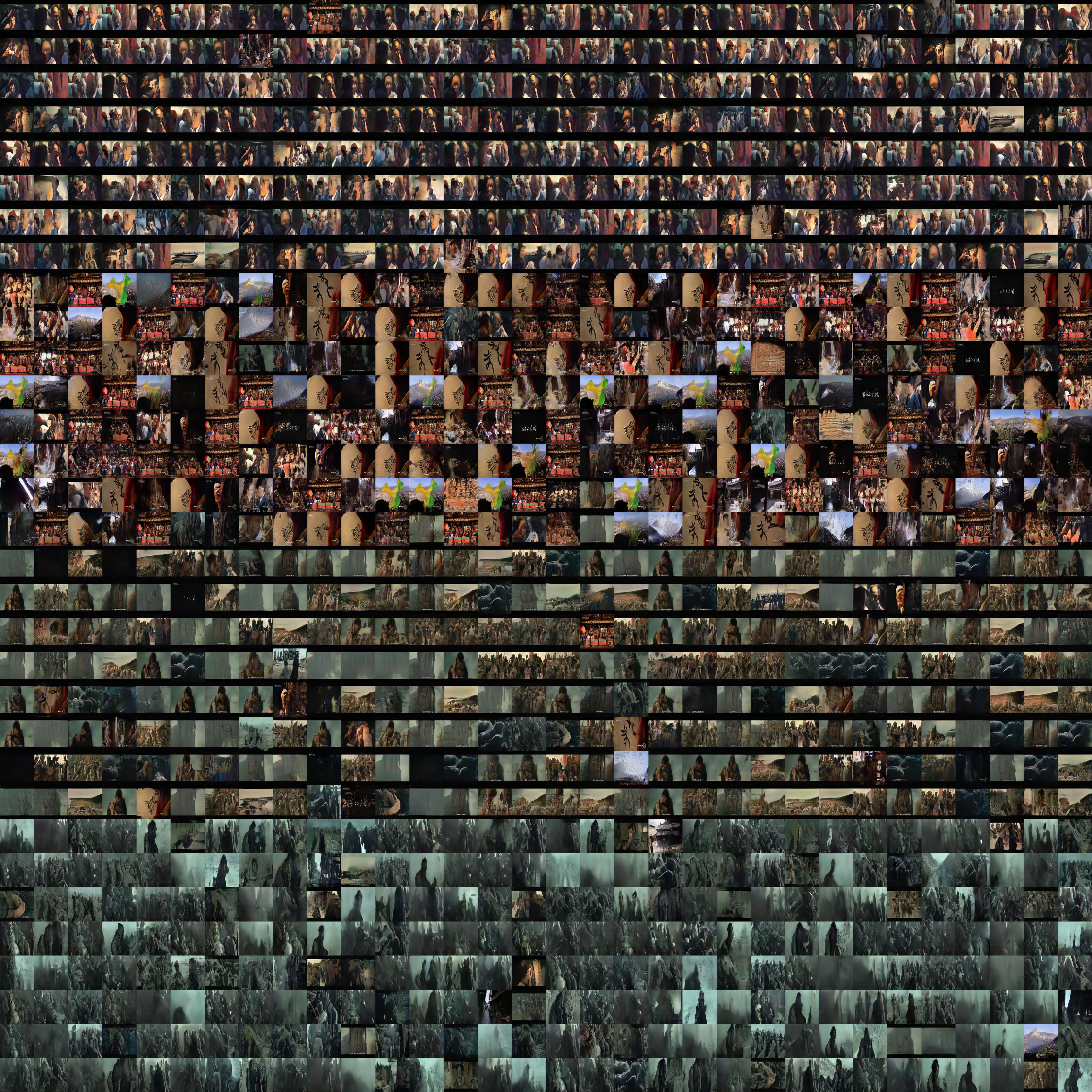}\vspace{1.0em}
    \caption{\textbf{Generated frames for training-set EEG}, SEED \citep{duan2013differential,zheng2015investigating}. Each tile is one frame generated from the EEG of a training segment.}
    \label{fig:additional_seed}
\end{figure*}

\begin{figure*}[tp]
    \centering
    \includegraphics[width=\linewidth]{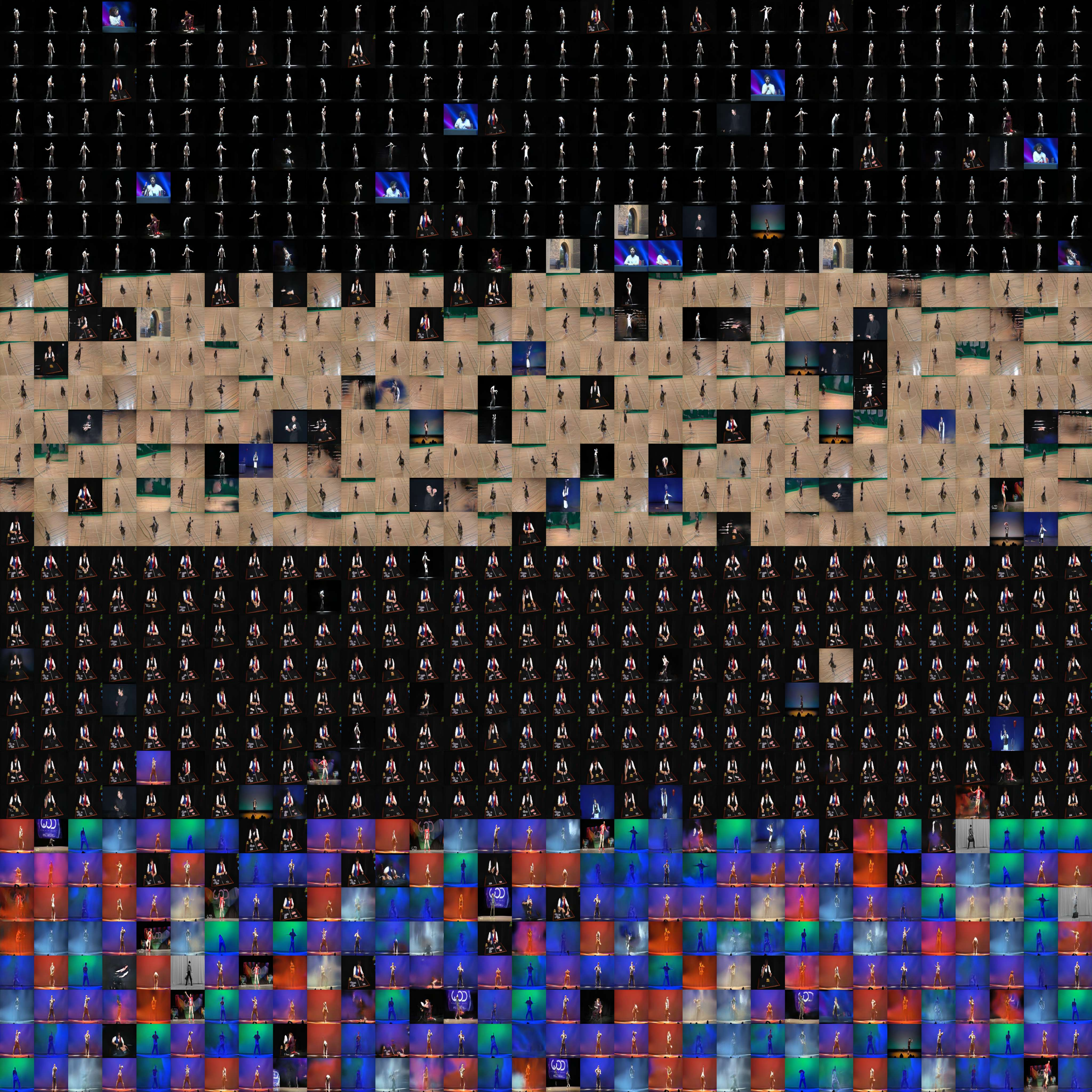}\vspace{1.0em}
    \caption{\textbf{Generated frames for training-set EEG}, EEG-Video Action \citep{yao2024identifying}. Each tile is one frame generated from the EEG of a training segment.}
    \label{fig:additional_action}
\end{figure*}

\begin{figure*}[tp]
    \centering
    \includegraphics[width=\linewidth]{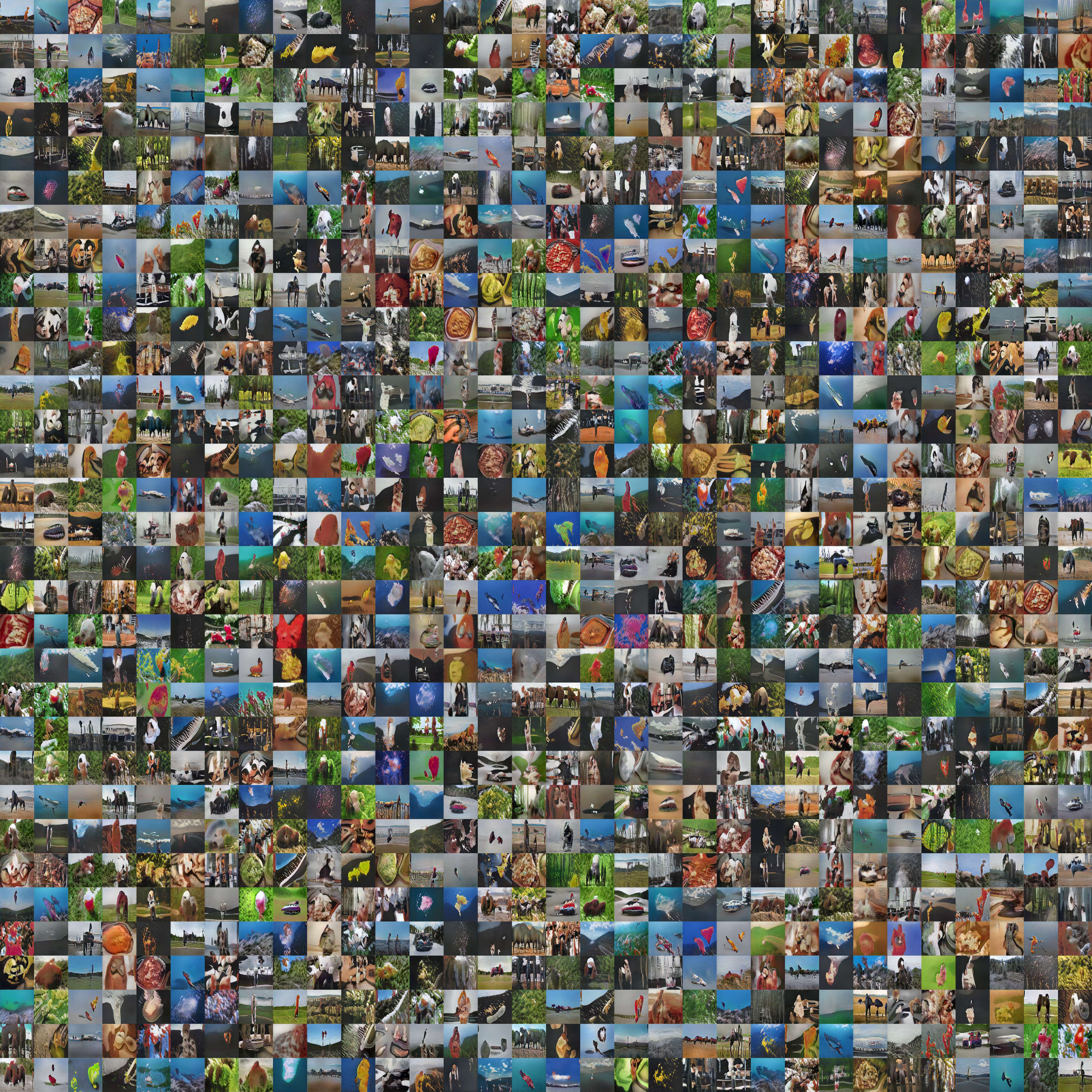}\vspace{1.0em}
    \caption{\textbf{Generated frames for training-set EEG}, SEED-DV \citep{eeg2video}. Each tile is one frame generated from the EEG of a training segment.}
    \label{fig:additional_seeddv}
\end{figure*}

Figures~\ref{fig:additional_seed}, \ref{fig:additional_action}, and~\ref{fig:additional_seeddv} show additional frames generated from \emph{training-set} EEG on SEED \citep{duan2013differential, zheng2015investigating}, EEG-Video Action \citep{yao2024identifying}, and SEED-DV \citep{eeg2video}. They illustrate the range of content the generator produces for each dataset. Test-set generations are shown in Figure~\ref{fig:teaser}(a) and evaluated in Table~\ref{table:quant_gan_eval}.

\end{document}